\documentclass[useAMS,usenatbib]{mn2e}
\usepackage{graphicx}
\usepackage{times}
\usepackage{amsmath}
\usepackage{amsfonts}
\usepackage{amssymb}
\usepackage{multirow}
\usepackage{longtable}
\RequirePackage{color}

\title[Contact binaries]{Origin of W UMa-type contact binaries -
age and orbital evolution}
\author[M. Y{\i}ld{\i}z]{M. Y{\i}ld{\i}z$^{}$\thanks{E-mail:
mutlu.yildiz@ege.edu.tr}\\
Department of Astronomy and Space Sciences, Ege University, Bornova, 35100 \.Izmir, Turkey}

\begin{document}
%\onecolumn
\twocolumn

\newcommand{\MS}{{\rm M}\ifmmode_{{\sun}}\else$_{{\sun}}$\fi}
\newcommand{\RS}{{\rm R}\ifmmode_{{\sun}}\else$_{{\sun}}$\fi}
\newcommand{\LS}{{\rm L}\ifmmode_{{\sun}}\else$_{{\sun}}$\fi}

\date{Accepted 2005 December 15. Received 2005 December 14; in original form 2005 October 11}

\pagerange{\pageref{firstpage}--\pageref{lastpage}} \pubyear{2006}

\maketitle

\label{firstpage}

\begin{abstract}
Recently, our understanding of the origin of W UMa-type contact binaries has become clearer.
Initial masses of their components were successfully estimated by Y{\i}ld{\i}z and Do\u{g}an using a new method 
mainly based on observational properties of overluminous secondary components. 
In this paper, we continue to discuss the results and make computations
for age and orbital evolution of these binaries.
It is shown that the secondary mass, according to its luminosity, also successfully predicts the observed radius. 
While the current mass of the primary component is determined by initial masses,
the current secondary mass is also a function of initial angular momentum.
We develop methods to compute the age of A- and W-subtype W UMa-type contact binaries { in terms of}  
initial masses and mass according to the luminosity of the secondaries.
Comparisons of our results with the mean ages from kinematic properties of these binaries and data pertaining to 
contact binaries in open and globular clusters, have increased our confidence on this method.
The mean ages of both A- and W-subtype contact binaries are found as 4.4  and 4.6 Gyr, respectively.
From kinematic studies, these ages are given as 4.5 and 4.4 Gyr, respectively. 
We also compute orbital properties of A-subtype contact binaries at the time of the 
first overflow. 
Initial angular momentum of these binaries is computed by comparing them with the well-known detached binaries.
The angular momentum loss rate derived in the present study for the detached phase is in very good agreement
with the semi-empirical rates available in the literature. 
In addition to the limitations on the initial masses of W UMa-type contact binaries, 
it is shown that the initial period of these binaries is less than about 4.45 d.

%The mass of primary components is determined by initial masses of components, in 90 per cent confident level.
%This shows that mass of the primary components changes less than 10 per cent, if it is not constant.

\end{abstract}


\section{Introduction}
The stars themselves are mysterious enough { to study}.  W UMa-type contact binaries (CBs) are, 
in many respects, excellent objects to research, being among the most observed stars and having experienced mass transfer between components.
The observational data of individual systems and statistical properties of entire data 
show so many regular and irregular variations that many contradictory ideas concerning their structure and evolution 
have been perpetuated. Recently, Y{\i}ld{\i}z \& Do\u{g}an (2013, hereafter Paper I) 
have discovered a new method to compute the initial masses of component stars from the observed masses and 
luminosity of the secondary component. 
This method gives the initial mass range for the secondaries as $M_{\rm 2i}$=1.3-2.6 \MS. That of 
the primaries is $M_{\rm 1i}=0.2-1.5 \MS$, which is in very good agreement with the required mass range for the angular momentum 
loss process via magnetic braking (Tutukov, Dremova \& Svechnikov  2004). Moreover, the initial masses of A- and W-subtypes have completely different
ranges for their secondaries: $M_{\rm 2i}> 1.8 \MS$ for the former and  $M_{\rm 2i}< 1.8 \MS$ for the latter.
In the present study, we further discuss some results of Paper I and develop new methods for estimation of age and angular 
momentum loss rates in detached and CB phases.

%Bir arastirma konusu olarak yildizlarin kendileri yeterince gizemlidir. 
%Bilesenleri arasinda kutle alisverisi gerceklesen ve en cok gozlenen yildizlar arasinda 
%bulunan W UMa tipi yakin ciftler ise bir cok bakimdan muhtesem bir arastirma konusdur. 
%Bu yildizlarin gerek tek tek gerekse 
%istatistiksel verileri o kadar duzenli ve duzensizlikler gosteriyor ki,
%bir birinden cok farkli ya da zit her turlu yaklasimin dogruluguna (ya da yanlisligina ) iliskin kanitlar sunan
%bundan daha uygun nesneler yoktur denebilir.
%Bu baglamda degerlendirebilecegimiz bir yaklasimla, YD2012 bu cift yildiz bilesenlerinin baslangic kutlelerini
%simdiki gozlemsel verilere (kutle ve isinimgucu) dayanarak hesaplamak icin ilk kez bir yontem kesfetti. 
%Buna gore, baslangicta agir kutleli olan
%simdiki ikinci bilesenlerin kutleleri $M_{\rm 2i}$ = 1.3 - 2.6 \MS, birinci bilesenlerin ise kutleleri $M_{\rm 1i}$= 0.2-1.5 \MS 
%araliginda yer almaktadir. Dahasi, baslangic kutleleri acisindan bakacak olursak A- ve W-alttipi yakin ciftler de 
%birbirinden ayrismaktadir: $M_{\rm 2i}>1.8$ ae 0.1 \MS ise A- alttipi, $M_{\rm 2i}<1.8$ ae 0.1 \MS ise W-alttipi yakin cift 
%olusmaktadir. Bu calismada ise, bu yildizlarin gerek nukleer gerekse yorunge evriminin anlasilmasi acisindan 
%yaslarinin hesaplanmasi ve bazi onemli sureclerin irdelenmesi amaclanmistir.

The formation history of A-subtype CBs is controlled in the pre-contact phase by two effects, namely,
nuclear evolution of the massive component ($M_{\rm 2i}$=1.8-2.7 M$_{\sun}$) and angular momentum evolution 
of the lighter component ($M_{\rm 1i}$=0.2-1.5 M$_{\sun}$).  
Depending on the initial masses and angular momentum reservoir, the detached phase ends up near 
the terminal-age of the main-sequence (TAMS) line in the Hertzsprung$-$Russel diagram (HRD).
{In the precursors of W-subtype CBs, however, both components experience efficient
angular momentum loss throughout or during part of their main-sequence MS phase of the massive
component. For some of their progenitors, angular momentum evolution is so fast that the components
may come into contact with each other before the massive component becomes a TAMS star.
}
%are effective in the angular momentum 
%loss course either throughout or in part of the MS phase of the massive component.
%For some of their progenitors, angular momentum evolution is so fast that the components 
%may stick to each other before the massive component becomes a TAMS star.
The maximum mass of their massive components is $1.8 \pm 0.1$ M$_{\sun}$. 
{ In addition,} for stars with $1.5 < M_{\rm 2i} < 1.8$ M$_{\sun}$ the secondary star is not effective 
in angular momentum loss process in the early detached phase of binary evolution, but becomes effective as its convective
zone deepens during evolution towards the TAMS line in the HRD.

For a typical A-subtype CB, the total initial mass is $\overline{M_{\rm Ti}}=\overline{M_{\rm 1i}}+\overline{M_{\rm 2i}}$= 3.0 M$_{\sun}$. 
The mean initial masses of the secondary 
and primary components are $\overline{M_{\rm 2i}}=2.0$ and  $\overline{M_{\rm 1i}}$=1.0 M$_{\sun}$, respectively. The mean total initial mass of W-subtype,
however, is $\overline{M_{\rm Ti}}$=2.5, 0.5 M$_{\sun}$ less than that of A-subtype CBs.
For W-subtype CBs, $\overline{M_{\rm 2i}}=1.7$ M$_{\sun}$ and $\overline{M_{\rm 1i}}= 0.8$ M$_{\sun}$.
%$\overline{M_{\rm 2i}}$ and $\overline{M_{\rm 1i}}$ for W-subtype CBs are 1.7 and  0.8 M$_{\sun}$, respectively.

Age determination of W UMa-type CBs is very important for a better understanding of binary evolution.
In addition to initial masses, if period ($P$) or  orbital dimension is estimated at the time of the first overflow
(FOF), for example, 
one can compute initial angular momentum and then deduce the rate of angular momentum loss as a first approximation.
A similar computation can also be made for the mass-loss. The present study also aims to make such computations.

Binaries have a variety of periods. If one of the components is a late-type star in the detached phase, then 
angular momentum loss causes the orbital period to decrease. If both components are late-type, then
angular momentum evolution is nearly twice as fast.
This may be the reason why W UMa-type CBs are observed in relatively young open clusters, such as  in Praesepe. 
Angular momentum evolution is one of the essential problems in stellar astrophysics 
and has been the subject of many papers on single or binary stars in the literature
(van 't Veer 1991; Eggleton, 2001; Demircan et al. 2006; St\c{e}pie\'{n}, 2006; Gazeas \& St\c{e}pie\'{n} 2008).
%(Give a short summary of the literature - St\c{e}pie\'{n}, Eggleton, Demircan etc.).  
%Unfortunately the suggested masses for the initial masses of W UMa-type CBs are very low  (Eggleton, 2001; St\c{e}pie\'{n} ...) 
%and therefore. angular momentum evolution is consider 
Although magnetic braking and tidal locking in binaries with late-type components are very complicated processes,
the angular momentum loss rates for such binaries derived by St\c{e}pie\'{n} (2006, see also Gazeas \& St\c{e}pie\'{n} 2008) and Demircan et al. (2006) are very easy to apply. We compare our results on angular momentum evolution of the detached precursors 
of W UMa-type CBs with these rates (see Section 4.2).

The final { end stage of evolution of W UMa-type CBs is also debated in the literature}
(see, e.g., Li et al. 2007; Eker, Demircan \& Bilir 2008).
The fate of a W UMa-type CB essentially depends on the ratio of angular momentum and mass-loss
during the contact phase.  
These systems seem to be the progenitors of blue stragglers observed in many clusters. 
Some of the blue stragglers are single stars and 
it seems reasonable to suppose that these were 
formed by merging 
components of W UMa-type CBs. In this case, the rate of angular momentum loss is so high that the components 
merge.
On the other hand, if mass-loss rate is very high, then the component stars remain separate.
The final form will then consist of either two brown dwarfs (Li et al. 2007) or two stars with very low mass. 
In the latter case, the age of the stars becomes infinitely long.
In order to predict the final products, 
further studies are required 
for mass and angular momentum loss rates in the W UMa-type CB phase. 
%In the present study, %Prior to this task, 
%age of these systems is found by a method based on use of their observed properties.

%Gaps in the MS of clusters:
%
%        Te   MZAMS  MTAMS  \\
%Gap   I  7200 1.5     1.8  \\
%Gap  II  6550 1.3     1.5  \\
%Gap III  5600     0.95     \\
%Gap  IV  4900     0.75     \\

The remainder of this paper is organized as follows. In Section 2, 
some basic results of Paper I are summarized and the $M_L-R$ relation for W UMa-type CBs is compared with
that of the well-known detached eclipsing binaries (DEBs).
Section 3 is devoted to the method for age computation of A- and W-subtype CBs and a comparison of results with
cluster member close binaries.
Orbital parameters of CBs and their evolution are considered and discussed in
Section 4.  Finally, our 
concluding remarks are given in Section 5.

\begin{figure*}
\includegraphics[bb=-25 340 290
480,width=80mm,angle=270]{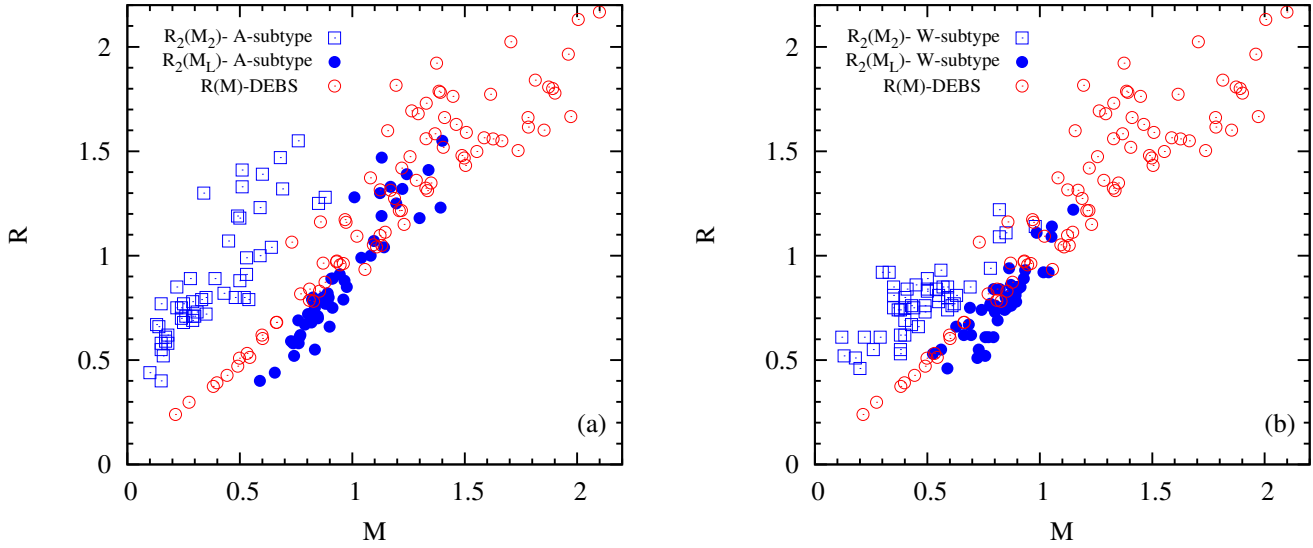}
\caption{The mass$-$radius relation for secondaries of the A- (a) and W-subtype (b) CBs (square). 
The filled circles represent mass according to luminosity-radius relation.
For comparison, the components of the well-known DEBs (circles) are also plotted.
Solar units are used. }
\end{figure*}
\section{The initial masses of W UMa-type CBs and connection with their structures}
%Typical binaries for A- and W-subtypes:
%
%A: M2i=2.0,  M1i=1.0, Mtop=3.0\\
%W  M2i=1.7,  M1i=0.8, Mtop=2.5

\subsection{Mass according to luminosity and radius relation}
{The secondary components of the W UMa-type CBs are over-luminous and over-sized according to their 
masses. In Paper I, Y{\i}ld{\i}z and Do\u{g}an consider this to be a result of the initial 
masses being much higher than the present (observed) masses.
They define $M_{L}$ of a secondary star as the mass according to its observed luminosity ($L_2$) by 
assuming $L_2=L_{\rm TAMS}$: $M_{L}=(L_2/1.5)^{0.237}$. 
They construct interior models with mass-loss by using the {\small MESA} stellar evolution code (Paxton et al. 2011)
and find a relation between the mass differences $\Delta M=M_{\rm 2i}-M_2$ and $\delta M=M_{L}-M_2$, where
$M_2$ is the present mass of the secondaries.
The expression derived from these models for initial mass of the secondary components 
in Paper I is a function of the mass difference
$\delta M$ and $M_2$:
% The expression they find for the initial mass of the secondary stars in terms of $\delta M$ and $M_{2}$ as 
%\begin{equation}
$M_{\rm 2i}=M_2+2.50 (\delta M-0.07)^{0.64}$,
%\end{equation}
where the masses are in units of solar mass.}

We want to test if $M_{L}$ is successful in predicting the present radii.
%$M_{L}$ is also 
%very suitable for our understanding of the secondary components observed radii ($R_2$).
$R_2$ of A- and W-subtypes are plotted with respect to $M_2$ %(square) 
and $M_{L}$ %(filled circle) 
in Figs. 1(a) and (b). 
We use the same data set as in Paper I.
The components of well-known DEBs %, circle) 
are also shown for comparison 
(Southworth, DEBCat:http://www.astro.keele.ac.uk/jkt/debcat/).
The secondary components are very large for their mass $M_2$. 
Hilditch, King \& McFarlane  (1988) have found that the radii 
of secondaries of the W-subtype CBs are 1.5 times larger than normal.
Although they have reported much larger factors for the secondaries of A-subtype CBs, we find that the difference between
the radii of the secondaries and normal stars is simply 0.5 \RS. For example,  a normal star of 0.4 \MS ~has a radius 
of 0.4 \RS ~(CU Cnc), but the radius of a secondary component in an A-subtype CB of the same mass (AQ Psc) is 0.9 \RS.
For W-subtype, there is no such single relation but a decreasing difference as mass increases.

The relation between $R_2$ and $M_{L}$, however,
is very close to the $M-R$ relation for the components of DEBs. 
This is a very good indicator for the fact that the entire structure of a secondary star with $M_{L}$ is very close to that of a single star
with mass $M_\star=M_{L}$. 
This result is somewhat surprising because one may expect that the oversized behaviour of a secondary star is mainly 
determined by the presence of a close companion. 

The slight differences between the $M_L-R$ relation for both A- and W-subtypes and the $M-R$ relation for DEBs are 
noticeable for low mass secondaries ($M_2 <0.15 \MS$, $M_L=0.5-0.8 \MS$). 
{This is most likely due to the influence of a very close 
companion ($P<0.38 $ d and $ a<2.7 \RS$) on the secondary radius.
}

%\begin{figure}
%\includegraphics[width=87mm,angle=270]{RvsMorML.W.ps}
%\caption{Same as Fig. 1, but for the secondaries of the W-subtype CBs. 
%}
%\end{figure}

%$M_{2i}$ and $M_{1i}$ are the masses in the detached phase. What are masses of components (M2p and M1p) when 
%system becomes contact?  

%The similarity of the secondaries' internal structure to that of a normal star with mass the same as 
%$M_{L}$ leads us to consider the evolution of these stars as being controlled by $M_{L}$.
%Internal structure of secondaries is very similar to a $normal$ star with mass the same as $M_{L}$. Therefore,
%we can deduce that evolution of these stars depends on $M_{L}$. 

{ In the light of these results we assume that the secondaries' internal structure is similar to
that of a star with the mass $M_L$.}
If these stars were isolated with their present features,
then their MS lifetime would be given by $M_{L}$ rather than $M_{2}$. 
This point has strong significance for our method to compute the ages of W UMa-type CBs (see Section 3).

\begin{figure*}
\includegraphics[bb=-25 340 290
480,width=80mm,angle=270]{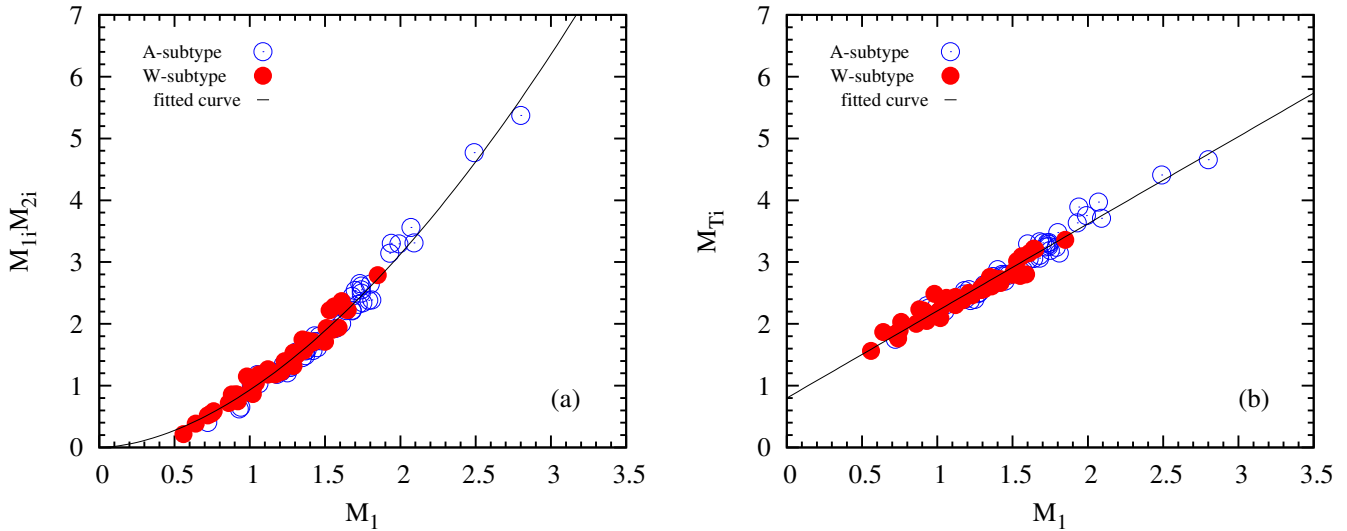}
\caption{Product (a) and summation (b) of initial masses of component stars as a function of the present 
mass of the primary component. The fitted curve is $(0.93\pm0.01) (M_1/{\rm M}_{\sun})^{1.75\pm0.02}$ in (a) and $(1.41\pm0.03) M_1/{\rm M}_{\sun}+(0.80\pm0.04)$ in (b).
}
\end{figure*}
%\begin{figure}
%\includegraphics[width=87mm,angle=270]{M1i+M2ivsM1.ps}
%\caption{Total initial mass of CBs as a function of present mass of primary component.
%}
%\end{figure}
%
\subsection{Correlation between initial and present masses}
It is well known that the gravitational force between two masses depends on the { product} of their masses and 
the inverse square
of the distance between them. We have the initial masses but not the distances between the component stars in their 
detached phase.  Therefore, we consider  the { product} of initial masses ($M_{\rm 1i} M_{\rm 2i}$) 
as a measure of tidal interaction. In Fig. 2(a) this 
{ product} for A- and W-subtypes is plotted with respect to the primary present mass $M_{1}$. 
%The A- and W-subtype CBs are represented by a circle and filled circle, respectively.
The solid line is for the curve fitted to the data. 
The present mass of the primary components is strongly dependent on their initial masses. The other 
parameters $-$ such as initial angular momentum, period and age of systems $-$  have  no role.    
In Fig. 2(b), the total initial mass is plotted against $M_1$.
There is a very clear linear relation between $M_{\rm Ti}$ and $M_1$. 
The fitting line is found as $M_{\rm Ti}=1.41 M_1+0.80$.
%This shows that initial angular momentum (or period) 
%is not important for $M_1$. 
$M_1$  can be computed from this equation as a function of initial masses (in solar mass): 
\begin{equation}
M_{\rm 1}=0.71 (M_{\rm 1i}+M_{\rm 2i})-0.57.
%M_{\rm 1c}=1.04 {(M_{\rm 1i}+M_{\rm 2i})^{0.57}}.
\end{equation}
The maximum difference between { the observed and computed $M_1$ from equation (1)}
is less than 10 per cent. This implies that
the mass of the primary  either remains constant or changes very little during the W UMa phase.

In Fig. 3, the total initial mass $M_{\rm Ti}$ is plotted against the present total mass $M_{\rm T}=M_{1}+ M_{2}$. 
The relation between $M_{\rm Ti}$ and $M_{\rm T}$ is also linear. 
The mean difference between $M_{\rm Ti}$ and $M_{\rm T}$ is 1.05 M$_{\sun}$ for A- %(circle) 
and 0.82 M$_{\sun}$ for W-subtype %(filled circle)
CBs. 
This shows that the A-subtype CBs have lost, on average, a mass 0.23 M$_{\sun}$ higher than W-subtype CBs.
The data are a bit more scattered in Fig. 3 than in Fig. 2(b). This implies that $M_2$ also 
depends on some other initial parameters as well as initial masses (see Section 4).

\section[]{Age of W UMa-type CB\lowercase{s}}
{
Age of a normal star can be found by some methods based on comparison of a computer
model and observational parameters. Time variation of stellar parameters allows such
methods. For a normal MS star with non-astreoseismic constraints, the highest (and
monotonic) changing (observable) parameters are luminosity and radius.
However, these parameters are a very sensitive function of mass, and composition. If
uncertainty in mass is high, then an accurate age is difficult to obtain. If the
uncertainty in mass is about 10 per cent, then age uncertainty is about 30 per cent.
In comparison, the effect of metallicity on age is greater. For example, a 1.5 \MS model
with solar composition (X=0.7024, Z=0.0172) has an MS lifetime of 2.1 Gyr. In
comparison when Z=0.0322, the age for a star of the same mass is 3.2 Gyr, a 50 per
cent longer MS lifetime. Therefore, any ages we find and estimate have some inherent
uncertainty.

On the other hand, even in the cases where we fit interior models to asteroseismic
constraints to find an age the agreement with other derived ages can be poor. For
alpha Centauri A and B, for example, ages from the non-asteroseismic and
asteroseismic constraints are found as about 9 Gyr and 5.6-5.9 Gyr (see, 
Miglio \& Montalban 2005; Y{\i}ld{\i}z 2007a),
respectively. Therefore, even with the best
state-of-the-art observational results there can still be some dispute or
uncertainty about the age of a star, this is compounded in the case of binary stars
where binary interactions will make it difficult to determine the true age. This is
in addition to the effect of uncertain current/initial stellar composition. Below, we
therefore aim to develop a method for determination of the age of observed W
UMa binaries in terms of the variable of stellar mass only to yield a rough, but
useful, stellar age.
}

{
As stated by Hilditch, King \& McFarlane (1988), the secondaries of A-subtype CBs are more evolved than that of W-subtype CBs.
Then, we anticipate that the detached phases of precursors of A- and W-subtype CBs are quite different.
%The Roche lobe of the secondary stars 
The  secondary stars at the end of this phase may overflow their Roche lobe for two very different reasons:\\

(i) due to rapid expansion 
%as a result of nuclear evolution 
after TAMS or 

(ii) due to the Roche lobe receding as a result of rapid orbital (angular momentum) evolution. \\

In case (i), the detached phase lasts about $t_{\rm MS}$ of the secondary star. 
This is simply the formation course of A-subtype. Since $M_{\rm 2i}>1.8\MS$ for A-subtype,
 the duration of the detached phase ($t_{\rm D}$) is less than $t_{\rm MS}(1.8\MS)$.
In case (ii), angular momentum loss rate is high because both components are
effective in the detached phase. This occurs in  the precursors of W-subtype CBs.
Their detached phase is shorter 
than the MS lifetime of their secondaries, $t_{\rm D} << t_{\rm MS}(M_{\rm 2i})$.
The shortest value of $t_{\rm D}$ for W-subtype with $M_{\rm 2i}<1.8\MS$ is $t_{\rm MS}(1.8\MS)/2$ (see below).  
} 

\subsection[]{Age of A-subtype CBs}
{
If $M_{\rm 2i}$ is higher than 1.8 $\MS$, then only the primary component is effective in the angular momentum 
loss process { via magnetic braking}. 
Such binaries become A-subtype. Since their angular momentum evolution is relatively slow, 
the Roche lobe overflow merely starts (i.e. mass transfer) after the secondary component completes its MS lifetime.
}

{ The mass transfer process obscures stellar quantities such as age.
Therefore, it seems to be very difficult to assign age to the W UMa-type 
CBs as they have three different evolutionary phases -
namely, detached, semidetached (SD) and CB phases.}
{Current age of a CB system ($t$) is the summation of durations of three phases.
They are detached ($t_{\rm D}$), SD ($t_{\rm SD}$) and contact phases ($t_{\rm CB}$). Then,
\begin{equation}
t_{\rm }=t_{\rm D}+t_{\rm SD}+t_{\rm CB},
\end{equation}
where $t_{\rm CB}$ is the time from the beginning of contact phase until present.}
As stated above, for A-subtype CBs, the time interval for the detached phase is very nearly the same as the 
MS lifetime of the secondary star: $t_{\rm D}\approx t_{\rm MS}(M_{\rm 2i})$. 
One can precisely compute $t_{\rm MS}$ as a function of $M_{\rm 2i}$. 
Y{\i}ld{\i}z (2013) derives an expression for $t_{\rm MS}$ from stellar { evolution} models as 
\begin{equation}
t_{\rm MS}=\frac{10}{(M/{\rm M}_{\sun})^{4.05}}(5.60\,\, 10^{-3}(\frac{M}{{\rm M}_{\sun}}+3.993)^{3.16}+0.042) ~{ \rm Gyr}.
%t_{\rm MS}/{ \rm yrs}=\frac{10^{10}}{(M/{\rm M}_{\sun})^{4.05}}(5.60\,\, 10^{-3}(\frac{M}{{\rm M}_{\sun}}+3.993)^{3.16}+0.042) ~{ \rm yrs}.
\end{equation}
%in units of year.
%The most explicit inference on the age of the W UMa-type CBs is that 
The detached phase of the precursor of an A-subtype system lasts slightly longer than the MS lifetime of its 
massive component [$t_{\rm MS}(M_{\rm 2i})$].  
%Age of an A-subtype system must be greater than MS lifetime of its massive component ($t_{MS}(M_{\rm 2i})$)
%in the detached phase. 
This time interval is very short for systems with high initial mass. 
On the other hand, the age of a system with a secondary of mass $M_L$ must be less than $t_{\rm MS}(M_L)$, or, 
more generally, $t_{\rm MS}(M_{\rm 2i}) < t_{\rm } < t_{\rm MS}(M_L)$ (see below).
From these constraints, we can derive an expression for $t_{\rm SD}+t_{\rm CB}$.

After the detached phase, nuclear and secular evolution develop under
control of each other. The SD phase starts with $M_{\rm 2i}$ and $M_{L}$ is 
the present presumptive mass according to luminosity.
The upper limit for these binaries, however, is the MS age of a star with a mass of $M=M_{L}$. The MS age of such
an isolated star is $t_{\rm MS}(M_{L})$. { In general, $ t_{\rm MS}(M_L) >> t_{\rm MS}(M_{\rm 2i})$.}
Perhaps, the most convenient age can be found by using the mean mass of the secondary components during 
binary evolution: 
\begin{equation}
\overline{M_2}=\frac{M_{\rm 2i}+M_{L}}{2}
\end{equation}
Then, we can take $t_{\rm SD}+t_{\rm CB}$ as
\begin{equation}
t_{\rm SD}+t_{\rm CB} \approx t_{\rm MS}(\overline{M_2}).
\end{equation}
{
If we insert $t_{\rm D}=t_{\rm MS}(M_{\rm 2i})$ and equation (5) in equation (2),
we find the age of an A-subtype CB:
\begin{equation}
t\approx t_{\rm MS}(M_{\rm 2i})+t_{\rm MS}(\overline{M_2}).
\end{equation}
}

{
The results of this simple method are listed in Table A1.
The basic properties of these binaries are given in Paper I.
%Age is given in the seventh column.
Although age of a CB is a very complicated problem and may be computed 
in a variety of ways, the present method gives reasonable results.
The average age for the A-subtype CBs is 4.37 Gyr. This value is, surprisingly, in very good agreement with
the age found for A-type CBs by Bilir et al. (2005), 4.48 Gyr. 
The results are listed in Table 1.}

In Table A1, uncertainties in ages of CBs are also listed. These uncertainties are derived from equation (6):
\begin{equation}
\Delta t=\frac{{\partial} t}{\partial M_2} \Delta M_2+\frac{\partial t}{\partial M_{\rm 2i}} \Delta M_{\rm 2i}.
\end{equation}
The uncertainty in $\Delta M_{\rm 2i}$ as a function of $\Delta M_{\rm 2}$ and $\Delta L_{\rm 2}$  is given in 
equation 11 of Paper I.
{The mean uncertainty given in Table 1 is the average of the uncertainties in age of individual binaries 
listed in Table A1.}
\begin{table}
\caption{Mean ages of A- and W-subtype CBs. The ages taken from Bilir et al. (2005) are the mean kinematical ages. 
}
\centering
\small\addtolength{\tabcolsep}{-3pt}
\begin{tabular}{lcc}
\hline
Type      &    Age (Gyr)  & Age (Gyr)\\
\hline
A-subtype $ ~~ ~~ $ & 4.37$\pm$1.23 & 4.48\\
W-subtype $ ~~ ~~ $ & 4.63$\pm$1.48 & 4.37\\
          &Present Study   $ ~~ $ & $ ~~ $ Bilir et al. (2005)  \\
\hline
\end{tabular}
\end{table}

\subsection[]{Age of W-subtype CB\lowercase{s}}
{ 
{If $M_{\rm 2i}$ is less than 1.8 $\MS$, then
both components have a convective envelope, and magnetic braking is  active in 
both stars. The angular momentum loss rate of these binaries is nearly twice as fast as  the precursors 
of the A-subtype CBs. Rapid angular momentum evolution causes the secondary component to fill its Roche lobe
before it completes its MS lifetime.
One should notice that a star of $1.8 \MS$ is an early-type star near the zero-age main-sequence (ZAMS) line and 
a convective envelope develops as it evolves towards TAMS. The outer convective structure of such a star is very similar 
to that of a star with $1.5 \MS$ near the ZAMS line.
That is to say, stars of $1.8 \MS$ have no role in magnetic braking near ZAMS 
but are very effective near TAMS. Another reason for the relatively thick convective envelope is high metallicity.
These might be the reason for why transition from A- to W-subtype occurs at $M=1.8 \MS$. }
The longest detached phase for A-subtype is $t_{\rm D}=t_{\rm MS}(1.8 \MS)=1.4 $ Gyr. 
Since the W-subtype CBs have two effective component stars, 
%As a result of two effective component stars in W-subtype CBs, 
the typical value of their $t_{\rm D}$ is $\approx t_{\rm MS}(1.8 \MS)/2$. 
For precursors of
typical W-subtype systems, the detached phase lasts about 0.7 Gyr. 
There are some such very young W UMa-type systems, for example, TX Cnc in Praesepe (see Table 2).
 
As discussed above, 
%the pre-contact evolutionary phases of the A- and W-subtype CBs are quite different from each other. 
%Progenitors of some W-subtype CBs angular momentum loss is nearly twice as fast as those  
%of the A-subtypes, at least in some part of their detached phase.
%Therefore, 
{the detached phase of the progenitors of W-subtype CBs takes less time than the 
MS lifetime of the component stars.}
This means that component stars touch each other before their massive component reaches the TAMS line.
For some systems, the angular momentum evolution is so fast that the detached phase is much shorter than 
the MS lifetime of the secondary components.
In such a case, the massive component is around the ZAMS line.
Therefore, equation (2) is not suitable for W-subtypes because the duration of the detached phase might be  
shorter than 
the MS lifetime. In this case, the right-hand side of equation (5) can be taken as the age of these systems:
}
\begin{equation}
t_{\rm } \approx t_{\rm MS}(\overline{M_2}).
\end{equation}
{
Equation (8) is valid if detached phase is negligibly small, i.e. $t_{\rm D}/t\approx 0$. 
If it is not so small, then we can take, for example,  $t_{\rm D}\approx t_{\rm MS}(M_{\rm 2i})/2$ and test which 
$t_{\rm D}$ is in better agreement with the observations than the other.

The average ages with negligible and non-negligible $t_{\rm D}$ are  
$4.63$ and $5.49$ Gyr, respectively. 
The result with $t_{\rm D}/t\approx 0$ is in very good agreement with the value found by 
Bilir et al. (2005), 4.37 Gyr. 
Therefore, the ages of the systems with $t_{\rm D}/t\approx 0$ are listed in Table A1.
The mean uncertainty in the mean age is found as $1.48$ Gyr. The mean results are also given in Table 1.  

}
\subsection[]{Comparison of A- and W-subtype of CBs and the cluster member close  binaries in period$-$age diagram}
{The ages we find for both A- and W-subtypes are in very good agreement with the kinematical 
ages given by Bilir et al. (2005).
We can also compare our findings with the cluster member close binaries in period$-$age diagram. 
%Bukowiecki et al. (2012) obtain the curve fitted to the data of these binaries in the period-age diagram. 
In Fig. 4, the periods of the A- %(circle) 
and the W-subtypes %(filled circle) 
are plotted with respect to their ages, together with the cluster member close binaries compiled by Bukowiecki et al. (2012).
The data of Bukowiecki et al. (2012) contain W UMa- and $\beta$-Lyrae-type close binaries. 
It is seen that there is no one-to-one relation between age and period. Any cluster has 
close binaries with very different periods.  
The solid line shows the curve fitted by Bukowiecki et al. (2012) for both type close binaries. 
Although the curve represents a tendency, its uncertainty is very high. 
The age$-$period relation of W-subtype has the same tendency as the curve derived by Bukowiecki et al. (2012). 
For A-subtype, however, the difference is significant for relatively young systems with $P>0.5$ d.
}
\begin{figure}
\includegraphics[width=77mm,angle=270]{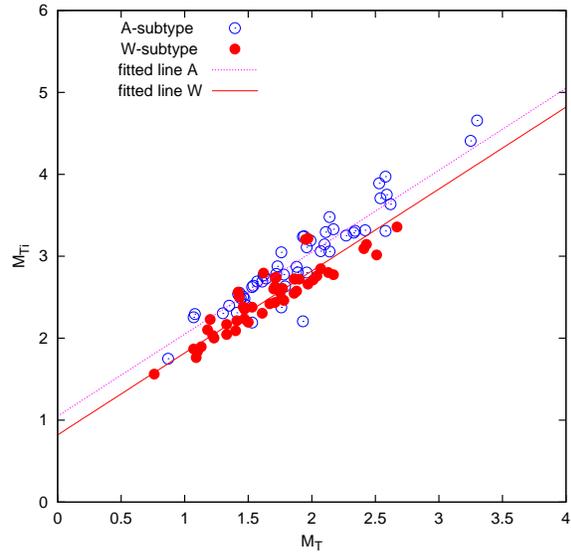}
\caption{Total initial mass of CBs as a function of total present mass. 
The fitted curves for A- and W-subtypes are $M_{\rm Ti}/M_{\sun}=
M_{\rm T}/M_{\sun}+(1.05\pm0.05)$ and $M_{\rm Ti}/M_{\sun}=M_{\rm T}/M_{\sun}+(0.82\pm0.02)$, respectively.
}
\end{figure}
\begin{figure}
%\includegraphics[width=77mm,angle=270]{Kiyas.P.vs.logt.ps}
%\includegraphics[width=77mm,angle=270]{Kiyas.P.vs.logt.Buko2012.ps}
%\caption{Age-period relation for the A- and W-subtype CBs.The dotted and solid lines are for the fitted curves derived by
%Bukowiecki et al. (2012) for W UMa-type CBs ($P({\rm d})=-0.09\log t({\rm y})+1.17$) and  for W UMa + $\beta$ Lyrae ($P({\rm d})=31.19/\log t({\rm y})-2.87$) type CBs, respectively.
\includegraphics[width=77mm,angle=270]{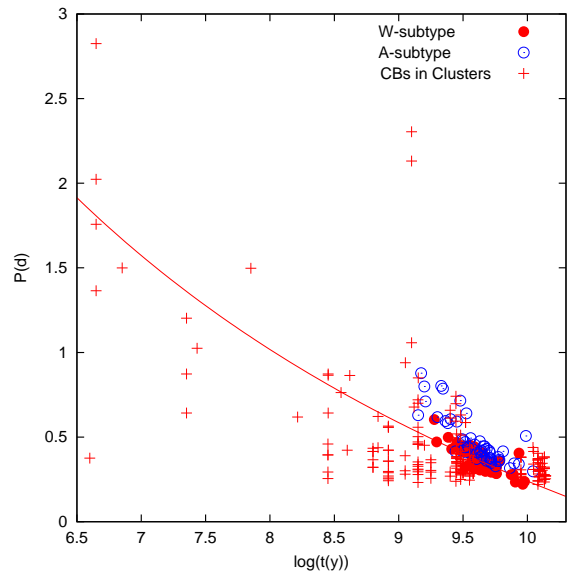}
\caption{Age$-$period relation for the A- and W-subtype CBs. The solid line is for the fitted curve derived by
Bukowiecki et al. (2012) for W UMa and  $\beta$ Lyrae ($P({\rm d})=31.19/\log t({\rm y})-2.87$) type CBs.
}
\end{figure}
The ages of W-subtype CBs from the fitting formula of Bukowiecki et al. (2012) are also listed in Table A1 ($t_{\rm B}$).
The mean difference between the ages from the fitting formula and from equation (6) or (8)
is 36 \%.
%  p 'out.A.tabTa+w.plot' u 16:(($18-$16)/($18)) ---- f g(x) 'out.A.tabTa+w.plot' u 16:(abs($18-$16)/($18)) via c

\subsection[]{Ages of TX Cnc, AH Vir, QX And and AH Cnc, and their clusters}
\begin{table}
\caption{Cluster member CBs.
}
\centering
\small\addtolength{\tabcolsep}{-3pt}
\begin{tabular}{lcccccccl}
\hline
Star   & $M_{\rm 1i}$ & $M_{\rm 2i}$ & $t_{\rm MS}$(Gyr)& $t$(Gyr) & $t_{\rm B}$(Gyr) & $t_{\rm cl}$(Gyr) & {St.} & Cluster \\
\hline
TX Cnc  & 0.50 & 1.70  & 1.6  &3.90 & 3.63 &  0.7  &  W  & Praesepe \\ % 0.50      -0.9
AH Vir  & 0.94 & 1.67  & 1.7  &3.25 & 3.08 &  3.0  &  W  & Wolf 630  \\% 0.94       1.3
QX And  & 0.62 & 1.92  & 1.1  &4.29 & 3.00 &  2.0  &  A  & NGC 752 \\   % 0.62      0.9
AH Cnc  & 0.93 & 1.85  & 1.3  &4.81 & 4.24 &  4.0  &  A  & M67\\       %   0.93     2.7
\hline
\end{tabular}
\end{table}
In our W UMa data sample, four binaries are cluster member. 
{The ages of these binaries are given in Table 2.
%The ages we found and predicted by the Bukowiecki et al. (2012) fitting formula are different 
%but have the same tendency
If we compare the ages of binaries with those of clusters, we notice that the cluster age is very low for TX Cnc.
This system is perhaps the youngest of the whole data set. 
The age we find for QX And, however, 
is twice the cluster age. For the other two binary systems, the age differences seem acceptable.
For AH Vir, for example, the differences between the three ages are less than 8 per cent.
Therefore, it should be stated that our method for age computation is better for A-subtype than for W-subtype. 
This is in contrast to the result obtained from comparison of our results with the data of the close binaries in clusters
(see Section 3.3).
The greatest difference appears for the youngest system TX Cnc.
Such exceptional binaries must either have very low angular momentum or 
very high angular momentum loss rate, or both.
However, we notice that the difference between the age of the present study and the cluster age depends on
$M_{\rm 1i}$: the smaller the value of $M_{\rm 1i}$, the greater is the difference ($t-t_{\rm cl}$).
% dt vs M_1i
% p 'Tab2.d' u 3:($6-$8)/$6 w lp pt 7 lt 3
}

\section[]{Estimation of orbital parameters at the first overflow}
\begin{figure}
\includegraphics[width=77mm,angle=270]{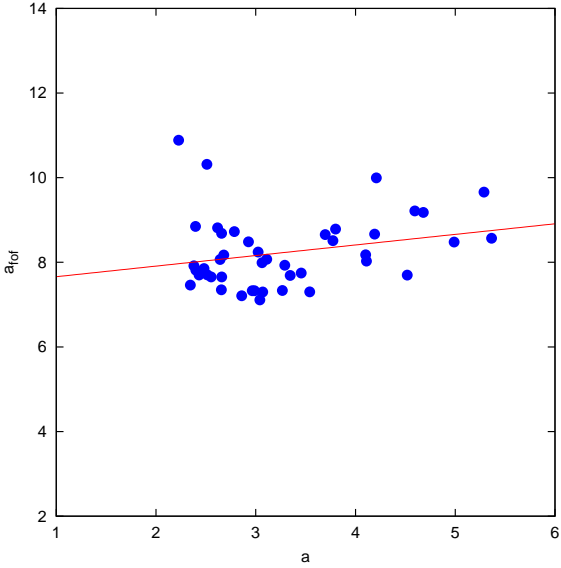}
\caption{ $a_{\rm fof}$ with respect to $a$, in units of solar radius.
$\overline{a_{\rm fof}}=8.1 \RS$. The fitted line is $a_{\rm fof}/\RS=(0.25\pm0.15)a/\RS+(7.41\pm0.49)$.
}
\end{figure}

\begin{figure}
\includegraphics[width=77mm,angle=270]{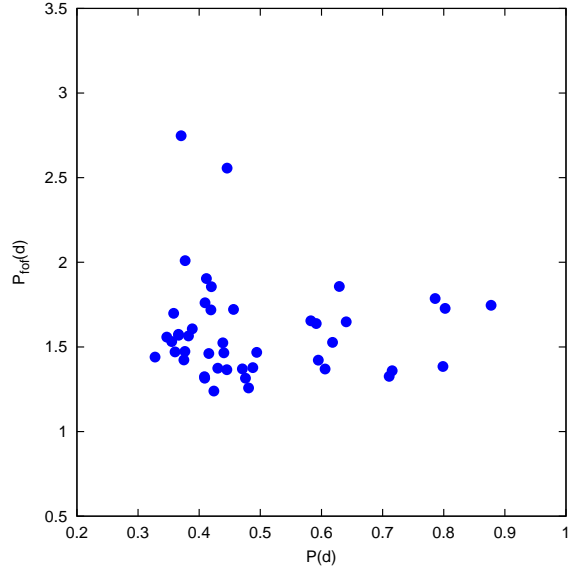}
\caption{$P_{\rm fof}$ with respect to $P$, in units of day.
$\overline{P_{\rm fof}}=1.6$ d.
}
\end{figure}
For the computation of  mean angular momentum and mass-loss rates in CBs, semimajor axis $a$ and $P$
are required at a certain time, in addition to the initial masses of the components. 
$a$ can be computed from the expression for the effective radius of Roche lobe 
given by Eggleton (1983). 
It seems reasonable to assume that, at least for A-subtypes, the mass transfer starts when the massive component of the detached phase 
fills its Roche lobe the first time. This point is known as the FOF. 
Then, the Roche lobe radius can be taken as $R_{\rm TAMS}(M_{\rm 2i})$.
The distance between the component stars at FOF is 
\begin{equation}
a_{\rm fof}=\frac{0.6q_{\rm i}^{-2/3}+\ln{(1+q_{\rm i}^{-1/3})}}{0.49q_{\rm i}^{-2/3}} {R_{\rm TAMS}}.
%\frac{r_{\rm A}}{a}=\frac{0.49q^{2/3}}{0.6q^{2/3}+\ln{(1+q^{1/3})}},\eqno(4)$$
\end{equation}
Note that $q_{\rm i}^{-1}=M_{\rm 1i}/M_{\rm 2i}$.
The values of $a_{\rm fof}$ found from equation (9) are plotted with respect to the present $a$ in Fig. 5. $a_{\rm fof}$ ranges from 7 to 11 $\RS$  and has a mean value of
$8.1 \RS$.  There is { a weak correlation} between $a_{\rm fof}$ and the present $a$.

Using Kepler's third law 
\begin{equation}
P_{\rm fof}=0.1159(a_{\rm fof}^3/(M_{\rm 1i}+M_{\rm 2i}))^{0.5},
%a=\frac{0.6q_{\rm i}^{2/3}+\ln{(1+q_{\rm i}^{1/3})}}{0.49q_{\rm i}^{2/3}} {r_{\rm TAMS}}
%\frac{r_{\rm A}}{a}=\frac{0.49q^{2/3}}{0.6q^{2/3}+\ln{(1+q^{1/3})}},\eqno(4)$$
\end{equation}
the period ($P_{\rm fof}$) at the end of the pre-contact phase is available. 
$P_{\rm fof}$ given in equation (10) is in units of days and plotted with respect to the present period $P$ in Fig. 6.
$P_{\rm fof}$ ranges from 1.24 to 2 d, except two A-subtype systems. The mean value of $P_{\rm fof}$ is about 
1.6 d. Although there is no correlation between $P_{\rm fof}$ and $P$ confirmed for the full range, there is an 
inverse relation for small values of $P$ ($0.5 < P$). 

$P_{\rm fof}$ and $a_{\rm fof}$ are given in Table A1. For W-subtypes, the values should be considered as 
just upper limits.

%Effective radii of the Roche lobes of both components are approximated
%to better than 1\% by formulae (Eggleton 1983)

%$$\frac{r_{\rm B}}{a}=\frac{0.49q^{-2/3}}{0.6q^{-2/3}+\ln{(1+
%q^{-1/3})}}.\eqno(5)$$

\subsection{Angular momentum evolution of W UMa-type CBs.}
Assuming components as point masses, the orbital angular momentum of a binary is given by
\begin{equation}
J=1.24\times 10^{52} (M_1+M_2)^{5/3} P^{1/3}\frac{q}{(1+q)^2} ~{ {\rm g~cm}^{-2}{\rm s}^{-1}}
\end{equation}
where masses are in units of slar masses and $P$ in units of day. 
%The initial angular momentum of W UMa-type CBs at the time of first overflow
%can also be computed using the initial masses and $P_{\rm fof}$ given in equation (8).
%
The angular momentum at the time of the FOF  ($J_{\rm fof}$) can be computed from equation (11) 
by using initial masses, and $P_{\rm fof}$. $\log(J_{\rm fof})$ is plotted with respect to the present angular momentum
$\log(J)$ in Fig. 7. There is a power-law relation between $J_{\rm fof}$ and $J$. The solid line is the fitting line. 
The slope in Fig. 7 is 0.46, which implies that $J_{\rm fof}\propto J^{0.46}$.
That is to say,  the present angular momentum depends on angular momentum at the time of the FOF:
\begin{equation}
J\propto J_{\rm fof}^{2.2}.
\end{equation}

For systems with the lowest angular momentum, $\Delta \log(J)=\log(J_{\rm fof}/J)\approx 1.0$. This means that
the present angular momentum of these systems is one tenth of $J_{\rm fof}$. For systems with the highest angular momentum, however,
$\Delta \log(J)=\log(J_{\rm fof}/J)\approx 0.4$. In this case, the present angular momentum is two fifths of 
$J_{\rm fof}$. 
\begin{figure}
\includegraphics[width=77mm,angle=270]{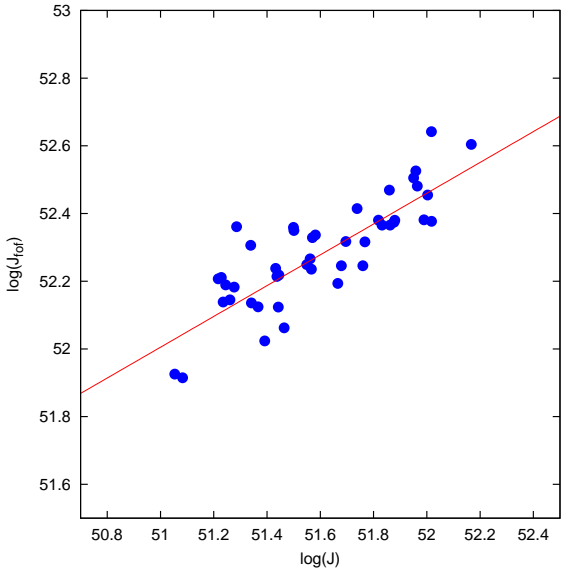}
\caption{$\log(J_{\rm fof})$ with respect to $\log(J)$, in cgs units. The fitted line is
$(0.46\pm0.05)\log(J)+(28.8\pm1.4)$
}
\end{figure}
\begin{figure}
\includegraphics[width=77mm,angle=270]{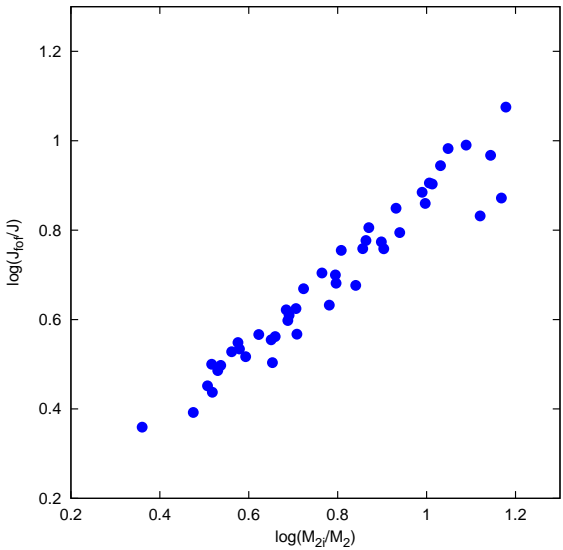}
\caption{Logarithmic change in angular momentum with respect to that of the secondary mass.
}
\end{figure}
In Fig. 8, logarithmic change in angular momentum is plotted with respect to logarithmic change in mass
of the secondaries.
It seems that the present mass of the secondary stars is determined by the amount of angular momentum loss after FOF.

%\begin{equation}
%\frac{\log(J_{\rm fof}/J)}{\log(M_{\rm i}/M_{\rm })}= 3.45
%\end{equation}
%where $M_{\rm si}$ and $M_{\rm s}$ are the initial and present masses of the system.
%The range is 2.5-5.0.
%
%A=M/J dJ/dM
%
%$1.6 < A <2.6$.
%
%This ratio is essentially the ratio of the SD phase.

\begin{figure}
\includegraphics[width=77mm,angle=270]{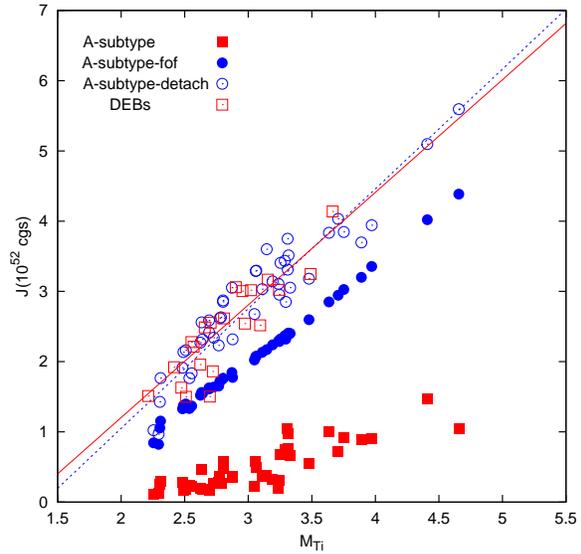}
\caption{Angular momentum with respect to initial mass for A-subtype 
W UMa-type CBs through their past evolutionary phases from 
detached to contact.  The squares are for the well-known eclipsing binaries (DEBs) having the same mass range as 
the initial masses of W UMa-type CBs.
The solid and dotted lines are for the fitted lines for DEBs and the estimated initial angular momentum of A-subtype CBs in the detached phase, respectively.
}
\end{figure}
\subsection{Angular momentum evolution in the contact phase}
Angular momentum loss rate in any phase of a binary system depends on its mass-loss rate and 
angular momentum content. In Fig. 9, angular momentum of A-subtype CBs 
at different evolutionary phases is plotted with respect to the total initial mass ($M_{\rm Ti}$).
The A-subtype CBs %are represented by filled squares. They 
have lost a large part of their 
angular momentum and therefore take place in the lower part of Fig. 9.
%The filled circles 
In Fig. 9, their angular momentum at the time of the FOF ($J_{\rm fof}$) is also plotted. 
On average, these binaries have lost 78 per cent of their $J_{\rm fof}$ (the range 55$-$90 per cent) during
the SD and contact phases; the larger the angular momentum is, the higher is the loss rate. 
Following several attempts, we derive a fitting formula for the angular momentum after FOF as a function of 
initial masses:
\begin{equation}
\frac{dJ}{dt}=-3.71\times 10^{41} M_{\rm 1i}^{0.82} M_{\rm 2i}^{4} ~~~~~~{ {\rm g~cm}^{-2}{\rm s}^{-1}{\rm y}^{-1}}, %(carpana dikkat )
\end{equation}
where masses are in solar masses.
%,  time is in years and angular momentum in cgs units.
Similarly, we derive another fitting formula for mass-loss:
\begin{equation}
\frac{dM}{dt}=-2.0\times 10^{-11}  M_{\rm 2i}^{4.38} ~~~~~~{ \MS {\rm y}^{-1}}%(carpana dikkat )
\end{equation}
\begin{figure}
\includegraphics[width=77mm,angle=270]{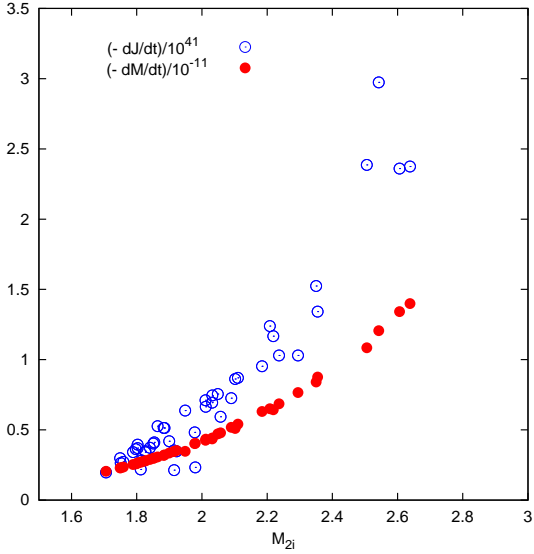}
\caption{Angular momentum (circle) and mass-loss (filled circle) rates with respect 
to initial mass of the secondaries. The units of angular momentum and mass-loss rates are 
${ {\rm g~cm}^{-2}{\rm s}^{-1}{\rm y}^{-1}}$ and ${ \MS {\rm y}^{-1}}$, respectively.
Note that dJ/dt and dM/dt are multiplied by $-1$.
}
\end{figure}
{The angular momentum and mass-loss rates are plotted with respect to $M_{\rm 2i}$ in Fig. 10. }
We note that these two formulae are very strong functions of $M_{\rm 2i}$ as luminosity.
This may imply that nuclear evolution of $M_{\rm 2i}$ dominates orbital evolution mainly driven by $M_{\rm 1i}$. 

Our present consideration does not allow for consideration of the SD and CB phases separately. 
The  angular momentum and mass-losses should be estimated 
in order to ascertain
the final product of W UMa-type CBs.
A study of near-CBs could help with this.

%From FOF to the present:
%dJ/dt= 0.0371 M1i**0.82*M2i**4
%
%dM/dt= 0.0200* M2i**4.38
%dP/dt= 0.44
%
%MdJ/dM/J=2.158

\subsection{Angular momentum evolution in the detached phase}
We compute typical angular momentum ($J_{\rm 0}$) in order to assess angular momentum evolution of precursors of A-subtype CBs in the detached phase.%(square). 
The angular momentum is a function of the total mass $M_{\rm Ti}$ in all phases. 
Angular momentum loss in the detached phase, however, depends on the convective structure of the primary 
components with a mass of $M_{\rm 1i}$.  Therefore, we assume that the loss rate ($dJ$/$dt$) is a function of $M_{\rm 1i}$:
\begin{equation}
\frac{dJ}{dt}=-c_J M_{\rm 1i}^{n},
\end{equation}
where $c_J$ and $n$ are free parameters to be determined from fitting $J_0$ to the angular momentum of the well-known 
DEBs ($J_{\rm DEB}$), 

\begin{equation}
J_{\rm fof}=J_0+\frac{dJ}{dt}\delta t= J_0+\frac{dJ}{dt}t_{\rm MS}.
\end{equation}
Then, we find the expression for $J_0$ in terms of $J_{\rm fof}$ and $M_{\rm 1i}$:
\begin{equation}
J_0= J_{\rm fof} - c_J M_{\rm 1i}^{n} t_{\rm MS}.
\end{equation}
The unit of $c_J$ is the same as the angular momentum.
If we take $M_{\rm 1i}$ and $t_{\rm MS}$ in units of solar mass and in units of year, respectively,
we find the value of $c_J$ and $n$ by fitting $J_0$ %(circle) 
to the angular momentum of DEBs %(square) 
as 
$c=0.8\times 10^{43}$ (cgs) and $n=1.5$:
\begin{equation}
\frac{dJ}{dt}=0.8\times 10^{43} M_{\rm 1i}^{1.5}.
\end{equation}
The dotted and solid lines in Fig. 9 represent the fitting lines for the DEBs and detached progenitors of W UMa binaries,
respectively.
They are very close to each other.
The progenitors of A-subtype CBs have lost 13$-$40 per cent of their initial angular momentum during the detached phase,
with a mean value of 29 per cent. However, most of their angular momentum is lost during the SD and CB phases. The 
total angular momentum lost by the progenitors of the A-subtype CBs' 
overall phases is about 84 per cent, i.e. the initial 
angular momentum of these binaries was six to seven times greater than their present angular momentum. 

The mean values of $a_{\rm 0}$ and $P_{\rm 0}$ are found from $J_{\rm 0}$ given in equation (17) as 16.3 \RS ~and 4.5 d, respectively. While $a_{\rm 0}$ ranges from  13.3 to 19.7 \RS, the interval for $P_{\rm 0}$ is from 2.8 to 6.3 d.
 $a_{\rm 0}$ and $P_{\rm 0}$ are given in Table A1.
 
%we take care angular momentum ($J_{\rm DEB}$) of corresponding DEBs (filled circle).

%P_ayrik= 2.8-6.3 d
%a_ayrik= 13.3-19.7 R_gun

%In Fig. 9, the filled circle represents angular momentum of DEBs (Southworth, 2012).  
\subsection{Comparison of angular momentum loss rates in the detached phase}
It is well known that the angular momentum loss rate for early-type stars is very slow in comparison with 
late-type stars.
Therefore, the rate we derive for the detached phase of A-subtype CBs is
only a function of the mass of their late-type component ($M_{\rm 1i}$). 
%\begin{equation}
%\frac{dJ}{dt}=-0.8\times 10^{43} M_1^{1.5}.
%\end{equation}
St\c{e}pie\'{n} (2006) gives the rate for close binaries as (see also Gazeas \& St\c{e}pie\'{n} 2008)
\begin{equation}
\frac{dJ}{dt}=-4.9\times 10^{41}(R_1^2 M_1+R_2^2 M_2)/P,
\end{equation}
where masses and radii are in solar units, period is in days, time in years and angular momentum in cgs units.
Equation (19) is a semi-empirical formula based on the angular momentum rate of single late-type stars.
The expression we derive is much easier than that of St\c{e}pie\'{n} (2006).
The mean ratio of the rate given in equation (18) to that of equation (19) for the detached phase is 0.63, which is 
satisfactory agreement. 
If we compute the initial angular momentum by going back from FOF, the initial angular momentum difference between the 
two rates is about 10 per cent.

Demircan et al. (2006) derive the time variation of angular momentum for detached chromospherically active binaries (CABs) as 
\begin{equation}
%J=J_0\exp({-3.48\times 10^{-10}t})=J_0 e^{-3.48\times 10^{-10}t}.
J=J_0 {\rm e}^{-3.48\times 10^{-10}t}.
\end{equation}
We also compare our results with that given by equation (20).
The mean ratio of the rate given in equation (18) to that of  equation (20) is 0.77 for the detached phase.
Many of the CABs studied by Demircan et al. are not in the initial mass range of W UMa-type CBs.
Despite this fact, the agreement between the two rates is very impressive.
The rate given in equation (18) is found from fitting the initial angular momentum of A-subtype to that of the 
well-known detached binaries with a period less than 5 d. In order to obtain better agreement between this 
rate and that of Demircan et al. (2006), the upper limit for the period should be reduced to 4.45 d. 
Then, $c_J=0.6 \times 10^{43}$ (cgs) for this upper limit.
This may imply that, in addition to the constraints on the initial masses of the components, 
the detached binaries with $P< 4.45$ d become A-subtype CBs.

\section{Conclusions}
{ 
In Paper I, a new method is developed for estimating initial masses of component stars in W UMa-type CBs,
based on introducing mass according to the luminosity of the secondaries.
In the present study, it is shown that 
an additional clue is the $M_L-R_2$ relation for secondary stars to support the idea that
the structure and evolution of these stars primarily depend on $M_L$: $M_L-R$ relation being in good agreement with
that of normal stars of detached binaries. 

Using the initial mass and mass according to luminosity, 
{ a new method is developed for 
age estimation of A- and W-subtypes. This is the first time that such a method of age estimation 
for W UMa-type CBs has been derived from the fundamental properties of secondary components.
We apply this method and find} that the mean ages of A- and W-subtypes are 
4.4 and 4.6 Gyr, respectively. These values are in very good agreement with the kinematic ages found
by Bilir et al. (2005).
{The ages we find are also compared with the ages of cluster member close binaries. 
The agreement is satisfactory for stellar age,
provided that the binary is not very young.

Our data sample contains four cluster member W UMa-type CBs; two A- and two W-subtypes.
The ages we find are in good agreement with the cluster ages, if $M_{\rm 1i}$ is about 1 \MS.
The difference is great for the CBs with $M_{\rm 1i}\approx 0.5-0.6$ \MS.
Determination of stellar parameters of such cluster member binaries is very important for 
our understanding of binary evolution.  
}
 
We show that there is a strong correlation between the present mass of primary ($M_{\rm 1}$) and the total
initial mass ($M_{\rm Ti}$). If we estimate ($M_{\rm 1}$) from this, { the difference 
between the observed and the estimated values of $M_{\rm 1}$ } is very small, being less than 10 per cent. 
This implies that $M_{\rm 1}$ either 
remains constant or changes very little during the W UMa phase, if at all.

The discovery of initial masses of W UMa-type CBs in terms of observed quantities leads us to
consider the binary evolution of these systems.
We compute the properties of A-subtype CBs at the time of the FOF 
by using  Eggleton's (1983) expression for the effective radius of Roche lobe.
%given by Eggleton (1983). 
The mean distance between the component stars in the former case is about 
$\overline{a_{\rm fof}}=8$ \RS ~and the mean period is 
$\overline{P_{\rm fof}}=1.6$ d. 
Computation of the period or mean distance enables us to also consider the angular momentum evolution of these systems.
We find that angular momentum at FOF is 2.5$-$10 times greater than the present angular momentum.
That is to say, these systems have lost a great deal of their angular momentum during the contact phase (SD+CB).

%we first compute initial angular momentum from $J_{\rm fof}$ by comparing 
%with angular momentum of the well known detached eclipsing binaries. 

For estimation of initial angular momentum in the detached phase, 
we compare $J_{\rm fof}$ with the angular momentum of the well-known DEBs.
Satisfactory agreement is obtained if
the rate of angular momentum loss in the detached phase is directly proportional to $M_{\rm 1i}^{1.5}$.
%
%We also estimate initial angular momentum of progenitors of A-subtype W UMa-type CBs by comparing them with 
%angular momentum of the well known detached eclipsing binaries. 
%
%and then initial parameters in the detached phase, 
The rate we derive 
is in very good agreement with the rates of St\c{e}pie\'{n} (2006) and Demircan et al. (2006).
We find that $\overline{a_{0}}=16.3$ \RS ~and $\overline{P_{0}}=4.5$ d.
In addition to the initial mass intervals for the components of W UMa-type CBs, we also obtain an upper 
limit for the initial period.
}

\section*{Acknowledgement}
I would like to thank Ros Elliot for her help in checking the language of the manuscript.
%The language of the manuscript is checked by Ros Elliot.
%The anonymous referee and Peter P. Eggleton are  acknowledged for
%their suggestions which improved presentation of the manuscript.

%J. Christensen-Dalsgaard is acknowledged for providing his adiabatic pulsation code. I thank also 
%Ay\c{s}e Lahur K{\i}rtun\c{c} for her suggestions which improved the language of the manuscript.
%This work is supported by The Scientific and 
%Technological Research Council of Turkey (T\"UB\.ITAK).
%Dr H. C. ... for a critical reading of the original version of the
%paper and an anonymous referee for very useful comments that improved
%the presentation of the paper.

\def \apj#1#2{ApJ,~{#1}, #2}
\def \aj#1#2{AJ,~{#1}, #2}
\def \astroa#1{astro-ph/~{#1}}
\def \pr#1#2{Phys.~Rev.,~{#1}, #2}
\def \prt#1#2{Phys.~Rep.,~{#1}, #2}
\def \rmp#1#2{Rev. Mod. Phys.,~{#1}, #2}
\def \pt#1#2{Phys.~Today.,~{#1}, #2}
\def \pra#1#2{Phys.~Rev.,~{ A}~{#1}, #2}
\def \asap#1#2{A\&A,~{#1}, #2}
\def \aandar#1#2{A\&AR,~{#1}, #2}
\def \apss#1#2{~Ap\&SS,~{{#1}}, #2}
\def \asaps#1#2{A\&AS,~{#1}, #2}
\def \arasap#1#2{ARA\&A,~{#1}, #2}
\def \pf#1#2{Phys.~~Fluids,~{#1}, #2}
\def \apjs#1#2{ApJS,~{#1}, #2}
\def \pasj#1#2{PASJ,~{#1}, #2}
\def \mnras#1#2{MNRAS,~{#1}, #2}
\def \ibvs#1{IBVS,~No.~{#1}}

\newpage
\onecolumn
\appendix
\section[]{Basic properties of W UMa  type binaries}
\small\addtolength{\tabcolsep}{-4pt}
%\begin{longtable}{@{}lrcccccccccclll@{}}
\begin{longtable}{lcccccccccccccc@{}}
\caption[Continued.]{Initial and present properties of the A- and W-subtype CBs.
The columns are organized as name, subtype, semimajor axis, period, 
primary and secondary masses, initial masses of primary and secondary before the mass transfer, 
mass of secondary according to luminosity,  
age (see Section 3), age from Bukowiecki et al. (2012),
semimajor axis and period at FOF, and initial semimajor axis and period.  
}\\
\hline
%   lclcccccr
%   l      & c&  l   &     c       &     c       &      c      &       c     &     c       &      c       &    & &\\
  Star     &Type &  $a$  &  $P$ & $M_{1}$ &$M_{2}$ &$M_{\rm 1,i}$&$M_{\rm 2,i}$&$M_{L}$& $t$ &$t_{\rm B}$ &$a_{\rm fof}$ &$P_{\rm fof}$& $a_{\rm D}$ &$P_{\rm D}$\\
           &  &(\RS)& (d)& (\MS)   &(\MS)   &    (\MS)    &    (\MS)    &(\MS)  & (Gyr)& (Gyr)&(\RS)         & (d)  &(\RS)         & (d)\\
\hline
\endfirsthead

\caption[-- continued from previous page]{-- continued from previous page}\\
\hline
  Star     &Type&  $a$ & $P$& $M_{1}$ &$M_{2}$ &$M_{\rm 1i}$&$M_{\rm 2i}$&$M_{L}$& $t$ &$t_{\rm B}$ &$a_{\rm fof}$ &$P_{\rm fof}$& $a_{\rm D}$ &$P_{\rm D}$\\
           &  &(\RS)& (d)& (\MS)   &(\MS)   &    (\MS)    &    (\MS)    &(\MS)  & (Gyr)& (Gyr)&(\RS)         & (d)  &(\RS)         & (d)\\

\hline
\endhead

\hline
\endfoot
HV      UMa& A& 4.988& 0.711& 2.80$\pm$ 0.60& 0.50$\pm$ 0.17& 2.11$\pm$ 0.57& 2.54$\pm$ 0.69& 1.30$\pm$ 0.12&  1.62$\pm$ 1.89&  0.48&  8.48&  1.33& 13.80&  2.75 \\
V376 And   & A& 5.364& 0.799& 2.49$\pm$ 0.06& 0.76$\pm$ 0.03& 1.90$\pm$ 0.10& 2.51$\pm$ 0.13& 1.40$\pm$ 0.02&  1.58$\pm$ 0.35&  0.30&  8.57&  1.38& 13.77&  2.82 \\
RR      Cen& A& 4.110& 0.606& 2.09$\pm$ 0.43& 0.45$\pm$ 0.10& 1.50$\pm$ 0.25& 2.21$\pm$ 0.38& 1.10$\pm$ 0.04&  2.53$\pm$ 1.77&  0.89&  8.03&  1.37& 15.04&  3.51 \\
V921 Her   & A& 5.288& 0.877& 2.07$\pm$ 0.05& 0.51$\pm$ 0.03& 1.37$\pm$ 0.05& 2.61$\pm$ 0.09& 1.34$\pm$ 0.00&  1.50$\pm$ 0.20&  0.20&  9.66&  1.75& 13.32&  2.83 \\
UZ  Leo    & A& 4.192& 0.618& 1.99$\pm$ 0.04& 0.60$\pm$ 0.02& 1.40$\pm$ 0.04& 2.35$\pm$ 0.07& 1.24$\pm$ 0.01&  2.00$\pm$ 0.24&  0.82&  8.67&  1.53& 13.99&  3.13 \\
V535 Ara   & A& 4.209& 0.629& 1.94$\pm$ 0.04& 0.59$\pm$ 0.02& 1.25$\pm$ 0.07& 2.64$\pm$ 0.15& 1.39$\pm$ 0.06&  1.41$\pm$ 0.39&  0.77& 10.00&  1.86& 13.34&  2.86 \\
AQ      Tuc& A& 4.102& 0.595& 1.93$\pm$ 0.21& 0.69$\pm$ 0.08& 1.42$\pm$ 0.20& 2.22$\pm$ 0.32& 1.22$\pm$ 0.03&  2.31$\pm$ 1.33&  0.95&  8.18&  1.42& 14.80&  3.46 \\
EF      Dra& A& 3.041& 0.424& 1.81$\pm$ 0.00& 0.29$\pm$ 0.00& 1.28$\pm$ 0.00& 1.86$\pm$ 0.00& 0.84$\pm$ 0.00&  4.63$\pm$ 0.00&  2.76&  7.11&  1.24& 19.56&  5.66 \\
V2388   Oph& A& 4.681& 0.802& 1.80$\pm$ 0.02& 0.34$\pm$ 0.01& 1.12$\pm$ 0.02& 2.35$\pm$ 0.04& 1.12$\pm$ 0.01&  2.14$\pm$ 0.16&  0.29&  9.18&  1.73& 13.77&  3.17 \\
AW      UMa& A& 3.024& 0.439& 1.79$\pm$ 0.14& 0.14$\pm$ 0.01& 1.13$\pm$ 0.02& 2.11$\pm$ 0.03& 0.90$\pm$ 0.00&  3.26$\pm$ 0.17&  2.51&  8.24&  1.52& 15.10&  3.78 \\
V776 Cas   & A& 3.063& 0.440& 1.75$\pm$ 0.04& 0.24$\pm$ 0.02& 1.14$\pm$ 0.05& 2.05$\pm$ 0.08& 0.91$\pm$ 0.01&  3.47$\pm$ 0.59&  2.48&  7.99&  1.47& 15.74&  4.05 \\
V592 Per   & A& 4.519& 0.716& 1.74$\pm$ 0.06& 0.68$\pm$ 0.04& 1.29$\pm$ 0.10& 2.03$\pm$ 0.15& 1.13$\pm$ 0.01&  3.01$\pm$ 0.88&  0.47&  7.70&  1.36& 16.41&  4.23 \\
XZ   Leo   & A& 3.456& 0.488& 1.74$\pm$ 0.05& 0.59$\pm$ 0.03& 1.26$\pm$ 0.06& 2.03$\pm$ 0.10& 1.08$\pm$ 0.00&  3.11$\pm$ 0.59&  1.83&  7.75&  1.38& 16.30&  4.21 \\
DK      Cyg& A& 3.345& 0.471& 1.74$\pm$ 0.00& 0.53$\pm$ 0.00& 1.24$\pm$ 0.00& 2.01$\pm$ 0.00& 1.04$\pm$ 0.00&  3.29$\pm$ 0.00&  2.04&  7.69&  1.37& 16.54&  4.32 \\
NN   Vir   & A& 3.540& 0.481& 1.73$\pm$ 0.02& 0.85$\pm$ 0.02& 1.36$\pm$ 0.06& 1.95$\pm$ 0.08& 1.20$\pm$ 0.01&  3.18$\pm$ 0.53&  1.91&  7.30&  1.26& 18.09&  4.90 \\
$\eta$  CrA& A& 3.697& 0.591& 1.72$\pm$ 0.04& 0.22$\pm$ 0.02& 1.06$\pm$ 0.03& 2.18$\pm$ 0.06& 0.98$\pm$ 0.00&  2.83$\pm$ 0.28&  0.97&  8.66&  1.64& 14.44&  3.53 \\
RZ      Tau& A& 3.111& 0.416& 1.70$\pm$ 0.16& 0.64$\pm$ 0.06& 1.21$\pm$ 0.14& 2.10$\pm$ 0.25& 1.14$\pm$ 0.03&  2.76$\pm$ 1.33&  2.92&  8.07&  1.46& 15.37&  3.84 \\
V401    Cyg& A& 3.800& 0.583& 1.68$\pm$ 0.00& 0.49$\pm$ 0.00& 1.09$\pm$ 0.00& 2.24$\pm$ 0.00& 1.13$\pm$ 0.00&  2.40$\pm$ 0.00&  1.02&  8.79&  1.65& 14.20&  3.40 \\
AQ   Psc   & A& 3.267& 0.476& 1.68$\pm$ 0.03& 0.39$\pm$ 0.02& 1.18$\pm$ 0.04& 1.89$\pm$ 0.07& 0.91$\pm$ 0.00&  4.27$\pm$ 0.59&  1.98&  7.33&  1.32& 18.45&  5.25 \\
AH      Aur& A& 3.290& 0.494& 1.68$\pm$ 0.05& 0.28$\pm$ 0.01& 1.10$\pm$ 0.02& 2.01$\pm$ 0.04& 0.91$\pm$ 0.01&  3.63$\pm$ 0.31&  1.76&  7.93&  1.47& 16.03&  4.22 \\
V839    Oph& A& 2.987& 0.409& 1.64$\pm$ 0.00& 0.50$\pm$ 0.00& 1.18$\pm$ 0.00& 1.88$\pm$ 0.00& 0.97$\pm$ 0.00&  4.08$\pm$ 0.00&  3.05&  7.33&  1.32& 18.49&  5.27 \\
FP   Boo   & A& 3.774& 0.640& 1.61$\pm$ 0.05& 0.15$\pm$ 0.02& 0.96$\pm$ 0.03& 2.09$\pm$ 0.06& 0.89$\pm$ 0.00&  3.36$\pm$ 0.39&  0.72&  8.51&  1.65& 14.90&  3.82 \\
V1073   Cyg& A& 4.595& 0.786& 1.60$\pm$ 0.02& 0.51$\pm$ 0.01& 1.00$\pm$ 0.01& 2.29$\pm$ 0.03& 1.17$\pm$ 0.00&  2.21$\pm$ 0.13&  0.32&  9.21&  1.79& 13.90&  3.31 \\
EX   Leo   & A& 2.859& 0.409& 1.57$\pm$ 0.03& 0.31$\pm$ 0.02& 1.07$\pm$ 0.05& 1.80$\pm$ 0.08& 0.83$\pm$ 0.01&  5.12$\pm$ 0.93&  3.06&  7.21&  1.32& 19.72&  5.99 \\
AP      Leo& A& 2.965& 0.430& 1.46$\pm$ 0.04& 0.43$\pm$ 0.02& 1.00$\pm$ 0.06& 1.80$\pm$ 0.11& 0.89$\pm$ 0.02&  4.83$\pm$ 1.18&  2.65&  7.33&  1.37& 19.23&  5.84 \\
FG      Hya& A& 2.344& 0.328& 1.45$\pm$ 0.03& 0.16$\pm$ 0.01& 0.90$\pm$ 0.02& 1.79$\pm$ 0.04& 0.74$\pm$ 0.00&  5.66$\pm$ 0.50&  5.31&  7.46&  1.44& 18.89&  5.80 \\
UX   Eri   & A& 3.070& 0.445& 1.43$\pm$ 0.03& 0.53$\pm$ 0.02& 1.00$\pm$ 0.05& 1.80$\pm$ 0.10& 0.94$\pm$ 0.01&  4.65$\pm$ 1.02&  2.40&  7.30&  1.37& 19.41&  5.92 \\
YY      CrB& A& 2.659& 0.377& 1.43$\pm$ 0.03& 0.35$\pm$ 0.01& 0.93$\pm$ 0.02& 1.85$\pm$ 0.05& 0.87$\pm$ 0.01&  4.61$\pm$ 0.46&  3.80&  7.65&  1.47& 17.73&  5.19 \\
CK      Boo& A& 2.452& 0.355& 1.42$\pm$ 0.00& 0.15$\pm$ 0.00& 0.85$\pm$ 0.00& 1.84$\pm$ 0.00& 0.76$\pm$ 0.00&  5.16$\pm$ 0.00&  4.40&  7.78&  1.53& 17.51&  5.17 \\
V566    Oph& A& 2.785& 0.410& 1.40$\pm$ 0.03& 0.33$\pm$ 0.01& 0.82$\pm$ 0.01& 2.06$\pm$ 0.03& 0.96$\pm$ 0.00&  3.31$\pm$ 0.21&  3.04&  8.72&  1.76& 14.87&  3.92 \\
TYC1174 a  & A& 2.642& 0.389& 1.38$\pm$ 0.01& 0.26$\pm$ 0.01& 0.83$\pm$ 0.02& 1.90$\pm$ 0.04& 0.85$\pm$ 0.01&  4.41$\pm$ 0.39&  3.50&  8.07&  1.61& 16.47&  4.69 \\
CN    Hyi  & A& 2.927& 0.456& 1.37$\pm$ 0.06& 0.25$\pm$ 0.02& 0.79$\pm$ 0.07& 1.98$\pm$ 0.16& 0.88$\pm$ 0.05&  3.88$\pm$ 1.48&  2.24&  8.48&  1.72& 15.45&  4.23 \\
GR      Vir& A& 2.399& 0.347& 1.37$\pm$ 0.16& 0.17$\pm$ 0.06& 0.81$\pm$ 0.10& 1.83$\pm$ 0.23& 0.76$\pm$ 0.02&  5.25$\pm$ 2.63&  4.65&  7.81&  1.56& 17.54&  5.24 \\
DZ   Psc   & A& 2.481& 0.366& 1.35$\pm$ 0.06& 0.18$\pm$ 0.02& 0.80$\pm$ 0.03& 1.83$\pm$ 0.08& 0.77$\pm$ 0.01&  5.22$\pm$ 0.87&  4.08&  7.85&  1.57& 17.45&  5.22 \\
HN   Uma   & A& 2.515& 0.383& 1.28$\pm$ 0.06& 0.18$\pm$ 0.01& 0.75$\pm$ 0.02& 1.76$\pm$ 0.06& 0.74$\pm$ 0.01&  5.93$\pm$ 0.83&  3.65&  7.70&  1.56& 18.46&  5.80 \\
V410 Aur   & A& 2.432& 0.366& 1.27$\pm$ 0.06& 0.17$\pm$ 0.03& 0.74$\pm$ 0.05& 1.75$\pm$ 0.11& 0.73$\pm$ 0.01&  6.06$\pm$ 1.52&  4.07&  7.70&  1.57& 18.56&  5.87 \\
SX      Crv& A& 2.160& 0.317& 1.25$\pm$ 0.04& 0.10$\pm$ 0.01& 0.72$\pm$ 0.02& 1.68$\pm$ 0.05& 0.66$\pm$ 0.01&  7.32$\pm$ 0.97&  5.75&  7.44&  1.52& 20.33&  6.86 \\
EQ      Tau& A& 2.481& 0.341& 1.22$\pm$ 0.03& 0.54$\pm$ 0.02& 0.91$\pm$ 0.06& 1.47$\pm$ 0.09& 0.82$\pm$ 0.01&  8.56$\pm$ 2.04&  4.84&  6.24&  1.17& 35.53& 15.92 \\
Y       Sex& A& 2.657& 0.420& 1.21$\pm$ 0.15& 0.22$\pm$ 0.03& 0.64$\pm$ 0.05& 1.91$\pm$ 0.15& 0.83$\pm$ 0.03&  4.38$\pm$ 1.44&  2.84&  8.68&  1.86& 15.62&  4.48 \\
V2357 Oph  & A& 2.670& 0.416& 1.19$\pm$ 0.01& 0.29$\pm$ 0.01& 0.72$\pm$ 0.02& 1.68$\pm$ 0.04& 0.76$\pm$ 0.00&  6.47$\pm$ 0.62&  2.92&  7.47&  1.53& 20.10&  6.74 \\
V404  Peg  & A& 2.679& 0.419& 1.18$\pm$ 0.00& 0.29$\pm$ 0.00& 0.67$\pm$ 0.00& 1.82$\pm$ 0.00& 0.82$\pm$ 0.00&  5.06$\pm$ 0.00&  2.85&  8.17&  1.72& 16.93&  5.12 \\
QX      And& A& 2.617& 0.412& 1.18$\pm$ 0.17& 0.24$\pm$ 0.04& 0.61$\pm$ 0.05& 1.92$\pm$ 0.17& 0.85$\pm$ 0.03&  4.29$\pm$ 1.58&  3.00&  8.81&  1.90& 15.48&  4.43 \\
VZ      Lib& A& 2.380& 0.358& 1.06$\pm$ 0.06& 0.35$\pm$ 0.03& 0.60$\pm$ 0.05& 1.71$\pm$ 0.13& 0.80$\pm$ 0.02&  6.00$\pm$ 1.89&  4.30&  7.92&  1.70& 18.51&  6.07 \\
OO      Aql& A& 3.329& 0.507& 1.05$\pm$ 0.02& 0.88$\pm$ 0.02& 0.91$\pm$ 0.23& 1.30$\pm$ 0.33& 1.01$\pm$ 0.05&  9.75$\pm$ 9.25&  1.62&  5.61&  1.04& 64.99& 40.89 \\
V508    Oph& A& 2.383& 0.345& 1.01$\pm$ 0.00& 0.52$\pm$ 0.00& 0.68$\pm$ 0.00& 1.51$\pm$ 0.00& 0.83$\pm$ 0.00&  7.90$\pm$ 0.00&  4.72&  6.91&  1.42& 26.47& 10.67 \\
DX      Tuc& A& 2.396& 0.377& 1.00$\pm$ 0.03& 0.30$\pm$ 0.01& 0.49$\pm$ 0.02& 1.81$\pm$ 0.06& 0.82$\pm$ 0.01&  5.06$\pm$ 0.67&  3.78&  8.85&  2.01& 16.16&  4.96 \\
TV      Mus& A& 2.510& 0.446& 0.94$\pm$ 0.14& 0.13$\pm$ 0.02& 0.34$\pm$ 0.02& 1.91$\pm$ 0.12& 0.79$\pm$ 0.03&  4.55$\pm$ 1.34&  2.40& 10.31&  2.56& 15.20&  4.58 \\
XY      Boo& A& 2.227& 0.371& 0.93$\pm$ 0.34& 0.15$\pm$ 0.05& 0.32$\pm$ 0.02& 1.98$\pm$ 0.15& 0.83$\pm$ 0.00&  4.03$\pm$ 1.13&  3.96& 10.89&  2.75& 15.07&  4.48 \\
TZ      Boo& A& 1.790& 0.298& 0.72$\pm$ 0.05& 0.15$\pm$ 0.04& 0.28$\pm$ 0.04& 1.47$\pm$ 0.19& 0.59$\pm$ 0.03& 11.05$\pm$ 5.80&  6.58&  8.57&  2.20& 20.56&  8.17 \\
AH      Cnc& A& 2.553& 0.360& 1.47$\pm$ 0.15& 0.25$\pm$ 0.03& 0.93$\pm$ 0.08& 1.85$\pm$ 0.16& 0.82$\pm$ 0.03&  4.81$\pm$ 1.78&  4.24&  7.65&  1.47& 17.70&  5.17 \\
RT      LMi& A& 2.656& 0.375& 1.31$\pm$ 0.05& 0.48$\pm$ 0.03& 0.88$\pm$ 0.07& 1.75$\pm$ 0.13& 0.90$\pm$ 0.01&  5.18$\pm$ 1.54&  3.84&  7.35&  1.42& 19.74&  6.27 \\

DN      Cam& W& 3.668& 0.498& 1.85$\pm$ 0.02& 0.82$\pm$ 0.02& 1.50$\pm$ 0.07& 1.86$\pm$ 0.09& 1.15$\pm$ 0.01&  1.26$\pm$ 0.37&  1.71&  6.82&  1.13& 13.81&  3.25 \\
V728    Her& W& 3.183& 0.471& 1.65$\pm$ 0.00& 0.30$\pm$ 0.00& 1.01$\pm$ 0.00& 2.19$\pm$ 0.00& 1.02$\pm$ 0.00&  1.80$\pm$ 0.00&  2.03&  8.78&  1.68& 13.80&  3.32 \\
V402    Aur& W& 3.766& 0.604& 1.64$\pm$ 0.05& 0.33$\pm$ 0.02& 1.01$\pm$ 0.03& 2.21$\pm$ 0.08& 1.04$\pm$ 0.01&  1.84$\pm$ 0.25&  0.90&  8.86&  1.70& 13.85&  3.33 \\
EF      Boo& W& 3.220& 0.430& 1.61$\pm$ 0.00& 0.82$\pm$ 0.00& 1.25$\pm$ 0.00& 1.90$\pm$ 0.00& 1.05$\pm$ 0.00&  1.95$\pm$ 0.00&  2.66&  7.27&  1.28& 13.33&  3.18 \\
ET    Leo  & W& 2.670& 0.347& 1.59$\pm$ 0.02& 0.54$\pm$ 0.01& 1.25$\pm$ 0.05& 1.55$\pm$ 0.05& 0.79$\pm$ 0.00&  2.03$\pm$ 0.58&  4.67&  6.07&  1.03& 12.71&  3.14 \\
AA      UMa& W& 3.400& 0.468& 1.56$\pm$ 0.00& 0.85$\pm$ 0.00& 1.21$\pm$ 0.00& 1.88$\pm$ 0.00& 0.99$\pm$ 0.00&  2.15$\pm$ 0.00&  2.07&  7.26&  1.29& 13.22&  3.17 \\
YY      Eri& W& 2.553& 0.321& 1.55$\pm$ 0.14& 0.62$\pm$ 0.07& 1.24$\pm$ 0.53& 1.53$\pm$ 0.45& 0.78$\pm$ 0.03&  2.20$\pm$ 5.57&  5.57&  6.01&  1.02& 12.66&  3.13 \\
ER      Ori& W& 3.224& 0.423& 1.53$\pm$ 0.00& 0.98$\pm$ 0.00& 1.27$\pm$ 0.00& 1.74$\pm$ 0.00& 1.05$\pm$ 0.00&  2.29$\pm$ 0.00&  2.77&  6.70&  1.16& 13.07&  3.15 \\
V502    Oph& W& 3.162& 0.453& 1.51$\pm$ 0.28& 0.56$\pm$ 0.08& 1.12$\pm$ 0.23& 1.73$\pm$ 0.36& 0.93$\pm$ 0.04&  2.39$\pm$ 2.57&  2.29&  6.88&  1.24& 12.69&  3.10 \\
V870    Ara& W& 2.681& 0.400& 1.50$\pm$ 0.01& 0.12$\pm$ 0.01& 0.91$\pm$ 0.02& 1.89$\pm$ 0.04& 0.77$\pm$ 0.00&  2.44$\pm$ 0.26&  3.25&  7.85&  1.52& 12.75&  3.15 \\
BB    Peg  & W& 2.676& 0.362& 1.42$\pm$ 0.02& 0.55$\pm$ 0.01& 1.07$\pm$ 0.03& 1.59$\pm$ 0.04& 0.87$\pm$ 0.00&  2.91$\pm$ 0.42&  4.21&  6.43&  1.16& 12.29&  3.06 \\
VY    Sex  & W& 3.014& 0.443& 1.42$\pm$ 0.02& 0.45$\pm$ 0.01& 0.99$\pm$ 0.02& 1.73$\pm$ 0.04& 0.87$\pm$ 0.00&  2.91$\pm$ 0.30&  2.43&  7.09&  1.33& 12.44&  3.08 \\
V417    Aql& W& 2.687& 0.370& 1.40$\pm$ 0.00& 0.50$\pm$ 0.00& 0.98$\pm$ 0.00& 1.74$\pm$ 0.00& 0.90$\pm$ 0.00&  3.05$\pm$ 0.00&  3.96&  7.11&  1.33& 12.44&  3.08 \\
AE      Phe& W& 2.698& 0.362& 1.38$\pm$ 0.06& 0.63$\pm$ 0.02& 1.02$\pm$ 0.14& 1.69$\pm$ 0.18& 0.87$\pm$ 0.03&  3.20$\pm$ 1.62&  4.18&  6.86&  1.26& 12.39&  3.07 \\
EZ      Hya& W& 2.958& 0.450& 1.37$\pm$ 0.00& 0.35$\pm$ 0.00& 0.85$\pm$ 0.00& 1.89$\pm$ 0.00& 0.89$\pm$ 0.00&  3.14$\pm$ 0.00&  2.34&  7.96&  1.57& 12.68&  3.16 \\
AH      Vir& W& 2.797& 0.407& 1.36$\pm$ 0.00& 0.41$\pm$ 0.00& 0.94$\pm$ 0.00& 1.67$\pm$ 0.00& 0.82$\pm$ 0.00&  3.35$\pm$ 0.00&  3.08&  6.95&  1.31& 12.18&  3.05 \\
V842    Her& W& 2.817& 0.419& 1.36$\pm$ 0.00& 0.35$\pm$ 0.00& 0.84$\pm$ 0.00& 1.90$\pm$ 0.00& 0.89$\pm$ 0.00&  3.08$\pm$ 0.00&  2.86&  8.04&  1.60& 12.70&  3.17 \\
UV      Lyn& W& 2.878& 0.415& 1.36$\pm$ 0.02& 0.50$\pm$ 0.01& 0.92$\pm$ 0.02& 1.80$\pm$ 0.05& 0.93$\pm$ 0.01&  3.32$\pm$ 0.32&  2.93&  7.47&  1.43& 12.51&  3.11 \\
W       UMa& W& 2.568& 0.334& 1.35$\pm$ 0.09& 0.69$\pm$ 0.05& 0.99$\pm$ 0.24& 1.77$\pm$ 0.32& 0.91$\pm$ 0.04&  3.44$\pm$ 2.23&  5.09&  7.24&  1.36& 12.54&  3.10 \\
QW      Gem& W& 2.557& 0.358& 1.31$\pm$ 0.04& 0.44$\pm$ 0.01& 0.90$\pm$ 0.03& 1.66$\pm$ 0.06& 0.83$\pm$ 0.01&  3.79$\pm$ 0.64&  4.31&  6.97&  1.33& 12.08&  3.04 \\
SS      Ari& W& 2.758& 0.406& 1.31$\pm$ 0.00& 0.40$\pm$ 0.00& 0.84$\pm$ 0.00& 1.78$\pm$ 0.00& 0.87$\pm$ 0.00&  3.65$\pm$ 0.00&  3.11&  7.57&  1.49& 12.35&  3.10 \\
V752    Cen& W& 2.588& 0.370& 1.30$\pm$ 0.00& 0.40$\pm$ 0.00& 0.84$\pm$ 0.00& 1.76$\pm$ 0.00& 0.85$\pm$ 0.00&  3.84$\pm$ 0.00&  3.97&  7.47&  1.47& 12.27&  3.09 \\
AM  Leo    & W& 2.656& 0.366& 1.29$\pm$ 0.01& 0.59$\pm$ 0.01& 0.94$\pm$ 0.03& 1.64$\pm$ 0.04& 0.90$\pm$ 0.00&  3.99$\pm$ 0.36&  4.09&  6.82&  1.29& 12.10&  3.04 \\
V781    Tau& W& 2.844& 0.408& 1.29$\pm$ 0.07& 0.57$\pm$ 0.03& 0.94$\pm$ 0.08& 1.61$\pm$ 0.13& 0.89$\pm$ 0.01&  3.99$\pm$ 1.19&  3.08&  6.70&  1.26& 12.03&  3.03 \\
V1191 Cyg  & W& 2.182& 0.313& 1.29$\pm$ 0.08& 0.13$\pm$ 0.01& 0.71$\pm$ 0.01& 1.85$\pm$ 0.03& 0.76$\pm$ 0.00&  3.85$\pm$ 0.21&  5.88&  8.19&  1.70& 12.44&  3.18 \\
RZ      Com& W& 2.476& 0.339& 1.23$\pm$ 0.09& 0.55$\pm$ 0.04& 0.89$\pm$ 0.12& 1.58$\pm$ 0.23& 0.89$\pm$ 0.03&  4.61$\pm$ 2.34&  4.93&  6.68&  1.28& 11.84&  3.01 \\
GM    Dra  & W& 2.303& 0.339& 1.21$\pm$ 0.04& 0.22$\pm$ 0.02& 0.67$\pm$ 0.02& 1.83$\pm$ 0.07& 0.79$\pm$ 0.00&  3.76$\pm$ 0.49&  4.92&  8.25&  1.74& 12.36&  3.19 \\
FU      Dra& W& 2.170& 0.307& 1.17$\pm$ 0.04& 0.29$\pm$ 0.01& 0.70$\pm$ 0.02& 1.68$\pm$ 0.05& 0.76$\pm$ 0.01&  4.81$\pm$ 0.51&  6.17&  7.51&  1.55& 11.86&  3.07 \\
U       Peg& W& 2.520& 0.375& 1.15$\pm$ 0.01& 0.38$\pm$ 0.01& 0.72$\pm$ 0.02& 1.66$\pm$ 0.05& 0.80$\pm$ 0.01&  4.70$\pm$ 0.50&  3.84&  7.37&  1.50& 11.81&  3.05 \\
AO      Cam& W& 2.354& 0.330& 1.12$\pm$ 0.01& 0.49$\pm$ 0.01& 0.77$\pm$ 0.02& 1.54$\pm$ 0.05& 0.80$\pm$ 0.00&  5.56$\pm$ 0.55&  5.24&  6.77&  1.34& 11.51&  2.98 \\
GZ      And& W& 2.279& 0.305& 1.12$\pm$ 0.02& 0.59$\pm$ 0.02& 0.75$\pm$ 0.06& 1.68$\pm$ 0.10& 0.84$\pm$ 0.01&  4.27$\pm$ 0.91&  6.24&  7.39&  1.49& 11.93&  3.06 \\
TW      Cet& W& 2.320& 0.317& 1.06$\pm$ 0.00& 0.61$\pm$ 0.00& 0.68$\pm$ 0.00& 1.74$\pm$ 0.00& 0.78$\pm$ 0.00&  4.32$\pm$ 0.00&  5.74&  7.82&  1.63& 12.04&  3.11 \\
BV      Dra& W& 2.376& 0.350& 1.04$\pm$ 0.02& 0.43$\pm$ 0.01& 0.59$\pm$ 0.02& 1.76$\pm$ 0.05& 0.87$\pm$ 0.01&  3.72$\pm$ 0.43&  4.55&  8.21&  1.78& 12.09&  3.17 \\
BX      Peg& W& 2.016& 0.280& 1.02$\pm$ 0.12& 0.38$\pm$ 0.05& 0.67$\pm$ 0.10& 1.42$\pm$ 0.20& 0.69$\pm$ 0.01&  7.76$\pm$ 3.46&  7.44&  6.55&  1.34& 11.02&  2.93 \\
OU      Ser& W& 1.989& 0.297& 1.02$\pm$ 0.01& 0.18$\pm$ 0.01& 0.50$\pm$ 0.02& 1.73$\pm$ 0.05& 0.72$\pm$ 0.01&  4.72$\pm$ 0.60&  6.62&  8.44&  1.90& 11.98&  3.22 \\
AB      And& W& 2.308& 0.332& 1.01$\pm$ 0.02& 0.49$\pm$ 0.01& 0.66$\pm$ 0.06& 1.54$\pm$ 0.14& 0.80$\pm$ 0.04&  5.53$\pm$ 2.00&  5.16&  7.05&  1.46& 11.35&  2.99 \\
SW      Lac& W& 2.380& 0.321& 0.98$\pm$ 0.00& 0.78$\pm$ 0.00& 0.61$\pm$ 0.00& 1.88$\pm$ 0.00& 0.86$\pm$ 0.00&  3.28$\pm$ 0.00&  5.59&  8.62&  1.86& 12.51&  3.25 \\
TY      Boo& W& 2.151& 0.317& 0.93$\pm$ 0.02& 0.40$\pm$ 0.01& 0.51$\pm$ 0.03& 1.66$\pm$ 0.10& 0.81$\pm$ 0.03&  4.58$\pm$ 1.12&  5.73&  8.11&  1.82& 11.71&  3.15 \\
VW      Cep& W& 1.972& 0.278& 0.93$\pm$ 0.02& 0.40$\pm$ 0.01& 0.57$\pm$ 0.05& 1.48$\pm$ 0.11& 0.66$\pm$ 0.02&  7.48$\pm$ 2.18&  7.56&  7.07&  1.52& 11.05&  2.98 \\
BW      Dra& W& 1.957& 0.292& 0.92$\pm$ 0.02& 0.26$\pm$ 0.01& 0.45$\pm$ 0.01& 1.65$\pm$ 0.05& 0.73$\pm$ 0.01&  5.23$\pm$ 0.59&  6.84&  8.29&  1.91& 11.69&  3.20 \\
TX      Cnc& W& 2.488& 0.383& 0.91$\pm$ 0.10& 0.50$\pm$ 0.06& 0.50$\pm$ 0.08& 1.71$\pm$ 0.29& 0.89$\pm$ 0.03&  3.90$\pm$ 2.32&  3.63&  8.32&  1.87& 11.90&  3.20 \\
V757    Cen& W& 2.345& 0.343& 0.88$\pm$ 0.00& 0.59$\pm$ 0.00& 0.49$\pm$ 0.00& 1.74$\pm$ 0.00& 0.85$\pm$ 0.00&  3.94$\pm$ 0.00&  4.77&  8.53&  1.93& 12.04&  3.24 \\
V829    Her& W& 2.273& 0.358& 0.86$\pm$ 0.02& 0.37$\pm$ 0.01& 0.47$\pm$ 0.02& 1.53$\pm$ 0.05& 0.74$\pm$ 0.01&  6.09$\pm$ 0.77&  4.31&  7.68&  1.74& 11.21&  3.07 \\
CC      Com& W& 1.587& 0.221& 0.72$\pm$ 0.02& 0.38$\pm$ 0.01& 0.35$\pm$ 0.07& 1.49$\pm$ 0.11& 0.53$\pm$ 0.01&  9.22$\pm$ 2.53& 11.52&  8.11&  1.97& 11.14&  3.18 \\
XY      Leo& W& 1.943& 0.284& 0.76$\pm$ 0.15& 0.46$\pm$ 0.06& 0.35$\pm$ 0.26& 1.68$\pm$ 0.46& 0.63$\pm$ 0.05&  5.75$\pm$ 6.06&  7.25&  9.07&  2.22& 12.03&  3.39 \\
V523    Cas& W& 1.662& 0.234& 0.75$\pm$ 0.03& 0.38$\pm$ 0.02& 0.36$\pm$ 0.07& 1.53$\pm$ 0.13& 0.56$\pm$ 0.01&  8.06$\pm$ 2.47& 10.48&  8.23&  1.99& 11.31&  3.20 \\
BH      Cas& W& 2.373& 0.406& 0.74$\pm$ 0.06& 0.35$\pm$ 0.03& 0.40$\pm$ 0.03& 1.36$\pm$ 0.12& 0.69$\pm$ 0.01&  8.61$\pm$ 2.46&  3.12&  7.23&  1.70& 10.56&  3.00 \\
RW      Dor& W& 1.866& 0.285& 0.64$\pm$ 0.00& 0.43$\pm$ 0.00& 0.24$\pm$ 0.00& 1.63$\pm$ 0.00& 0.68$\pm$ 0.00&  5.71$\pm$ 0.00&  7.17&  9.84&  2.62& 12.30&  3.66 \\
RW      Com& W& 1.471& 0.237& 0.56$\pm$ 0.06& 0.20$\pm$ 0.03& 0.15$\pm$ 0.01& 1.41$\pm$ 0.12& 0.59$\pm$ 0.01&  9.49$\pm$ 2.80& 10.20&  9.68&  2.79& 11.77&  3.74 \\
\hline
\end{longtable}
$^a$) TYC 1174-344-1
\twocolumn

\label{lastpage}

\end{document}